\documentclass[12pt]{article}

\usepackage{ulem}
\usepackage{amsmath}
\usepackage{latexsym}
\usepackage{graphicx}
\usepackage{amstext}
\usepackage{amssymb}
\usepackage{graphics}
\usepackage{color}
\usepackage{algorithm}
\usepackage{authblk}
\usepackage{algpseudocode}
\newcommand{\mb}[1]{\mbox{\boldmath$#1$}}
\newcommand{\p}{\partial}
\newcommand{\ds}{\displaystyle}
\newcommand{\beq}{\begin{eqnarray}}
\newcommand{\beqq}{\begin{eqnarray*}}
\newcommand{\eeq}{\end{eqnarray}}
\newcommand{\eeqq}{\end{eqnarray*}}
\newcommand{\eps}{\varepsilon}
\newcommand{\x}{\mbox{\boldmath$x$}}

\newcommand{\s}{\mbox{\boldmath$s$}}

\newcommand{\w}{\mbox{\boldmath$w$}}

\newcommand{\Bb}{\mbox{\boldmath$b$}}

\font\bb=msbm10 at 12pt 

\def\rR{\hbox{\bb R}}
\def\nN{\hbox{\bb N}}

\def\pP{\hbox{\bb P}}
\def\eE{\hbox{\bb E}}
\def\ds#1{\displaystyle{#1}}

\definecolor{red}{rgb}{1,0,0}

\DeclareMathOperator\erfc{erfc}
\DeclareMathOperator\erf{erf}

\DeclareMathOperator\Var{Var}
\DeclareMathOperator\FWHM{FWHM}
\begin{document}
\title{Modeling, Segmenting and Statistics of Transient Spindles via Two-Dimensional Ornstein-Uhlenbeck Dynamics}
\author[1]{C. Sun}
\author[1]{D. Fettahoglu}
\author[1]{D. Holcman}
\affil[1]{{\small Group of Data Modeling, Computational Biology and Predictive Medicine, \'Ecole Normale Sup\'erieure, Universit\'e PSL, Paris, France.}}
\date{\today}
\maketitle
\begin{abstract}
We develop here a stochastic framework for modeling and segmenting transient spindle-like oscillatory bursts in electroencephalogram (EEG) signals. At the modeling level, individual spindles are represented as path realizations of a two-dimensional Ornstein–Uhlenbeck (OU) process with a stable focus, providing a low-dimensional stochastic dynamical system whose trajectories reproduce key morphological features of spindles, including their characteristic rise–decay amplitude envelopes. On the signal processing side, we propose a segmentation procedure based on Empirical Mode Decomposition (EMD) combined with the detection of a central extremum, which isolates single spindle events and yields a collection of oscillatory atoms. This construction enables a systematic statistical analysis of spindle features: we derive empirical laws for the distributions of amplitudes, inter-spindle intervals, and rise/decay durations, and show that these exhibit exponential tails consistent with the underlying OU dynamics. We further extend the model to a pair of weakly coupled OU processes with distinct natural frequencies, generating a stochastic mixture of slow, fast, and mixed spindles in random temporal order. The resulting framework provides a data-driven framework for the analysis of transient oscillations in EEG and, more generally, in nonstationary time series.
\end{abstract}
\paragraph{{\bf Abbreviations:}} EEG = ElectroEncephaloGram; GA = General Anesthesia; OU = Ornstein-Uhlenbeck; $\alpha$S = Alpha-Suppression;
\section{Introduction}
Studying random oscillatory signals in systems lacking a well-defined analytical structure poses significant theoretical and practical challenges. Classical tools such as Fourier transforms provide global spectral information, but fail to capture localized and transient phenomena. In contrast, wavelet transforms and related multiscale methods have been successfully used to detect local bursts, heterogeneity, and pointwise irregularities in time-series data \cite{flandrin1992wavelet,jaffard2001wavelets}. However, beyond descriptive decomposition, understanding the emergence and variability of transient patterns in systems driven by stochastic dynamics requires the development of generative models and estimation of underlying parameters.

Electroencephalogram (EEG) signals exemplify this setting. They arise from the collective dynamics of excitatory and inhibitory neuronal populations, generating rhythmic voltage fluctuations through bursting and synaptic interactions. These fluctuations manifest as transient events across multiple frequency bands. Notably, alpha-band spindles (8--12 Hz) are short-lived oscillatory bursts that occur in various physiological states, including quiet wakefulness, early sleep, general anesthesia, meditation, and sensory-evoked activity \cite{buzsaki2006rhythms,da1973organization,bacsar2012short,bacsar2012review,bacsar2012brain}.

Stochastic differential equations (SDEs) have been widely employed to model network-level neuronal dynamics \cite{destexhe2000,holcman2006emergence}. In particular, such models can capture stochastic transitions between multiple attractors, often separated by a limit cycle, effectively reproducing transient oscillatory behavior and metastability \cite{duc2014oscillatory,daoduc2016,zonca2021escape,zonca2021modeling}. These systems not only account for burst-like oscillations but also reveal, through two-dimensional phase-space analysis, how the proximity of attractors to the boundary of a limit cycle induces non-Poissonian residence-time distributions \cite{holcman2006emergence,daoduc2015,daoduc2016}.

Linearization of the dynamics in the vicinity of a stable focus attractor yields a two-dimensional Ornstein-Uhlenbeck (OU) process. Under specific parameter regimes, this linear system exhibits damped oscillations with a characteristic phase delay relative to the original nonlinear limit-cycle dynamics—an effect that can be computed analytically \cite{verechtchaguina2006_1,verechtchaguina2006_2,thomas2019phase,perez2021isostables}.

In contrast to this classical interpretation, the present study investigates a complementary regime in which the two-dimensional OU process, driven by stochastic forcing, gives rise to recurrent bursts of oscillatory activity—akin to EEG spindles—without the need for a nearby deterministic cycle. We show that such random bursts arise spontaneously and recurrently, centered around a characteristic frequency, as a manifestation of resonance in a linear system with rotational drift \cite{schuss2009diffusion}. This mechanism provides a minimal yet effective model for generating transient oscillations with statistically quantifiable structure.

The remainder of this paper is organized as follows. In Section 2, we present the classical properties of the two-dimensional Ornstein-Uhlenbeck (OU) process and highlight its capacity to generate oscillatory trajectories under stochastic forcing. Section 3 introduces a  mechanism for spindle generation and presents a robust segmentation algorithm based on Empirical Mode Decomposition (EMD). This algorithm identifies spindle events as intervals between two successive local minima enclosing a single maximum, enabling reliable extraction of transient oscillations \cite{jaffard2001wavelets,flandrin2004empirical}. In Section 4, we apply this segmentation to compute key statistical descriptors of spindles, including their duration, peak amplitude, and inter-spindle interval distributions. Section 5 focuses on characterizing the temporal asymmetry of spindles by computing the statistics of their ascending and descending phases using first-passage time analysis. In Section 6, we extend the model to a system of two weakly coupled OU processes with distinct natural frequencies. This coupling leads to the emergence of diverse spindle types—slow, fast, and mixed—appearing in a stochastic sequence without discernible temporal patterns. We conclude in Section 7 with a discussion of the implications of this modeling framework for understanding transient oscillatory activity in EEG recordings, and its potential for broader applications in stochastic signal analysis.
\section{Two-dimensional Ornstein-Uhlenbeck process to model spindle dynamics}\label{sec:statistics}
In this section, we introduce a model for local transient bursts based on two-dimensional Ornstein-Uhlenbeck (OU) processes (Fig.~\ref{fig:1}). These transient oscillatory events, referred to as \emph{spindles}, emerge as manifestations of a resonance phenomenon induced by stochastic forcing. As we shall demonstrate, the interplay between rotational drift and noise in the OU dynamics gives rise to spontaneous bursts oscillating around a characteristic frequency—a behavior closely related to stochastic resonance effects \cite{lindner2004,verechtchaguina2004spectra}.
\subsection{Two-dimensional anti-symmetric Ornstein-Uhlenbeck process generates spindle dynamics}
\begin{figure*}
\centering
\includegraphics[width=0.8\linewidth]{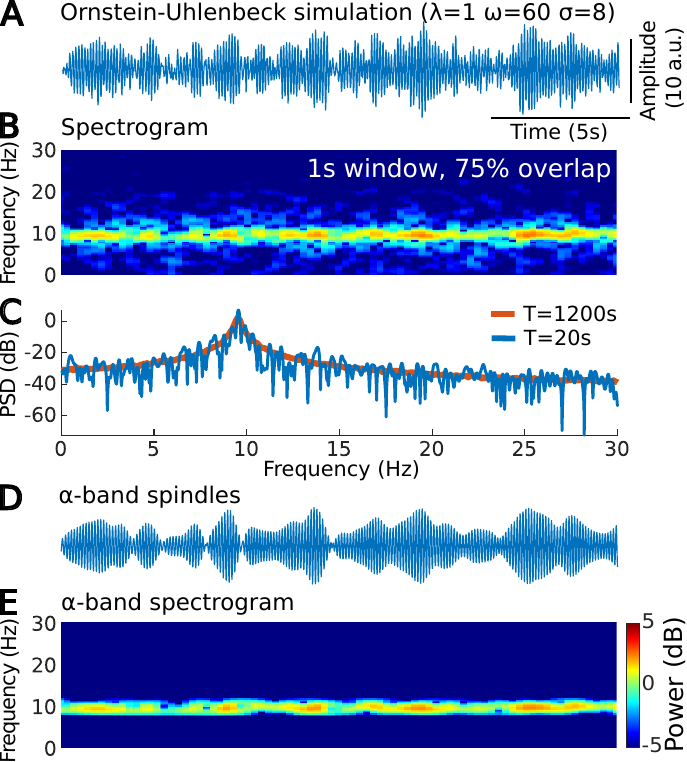}
\caption{{\bf Spectral decomposition reveals spindles dynamics in an Ornstein-Uhlenbeck simulation.}
{\small
{\bf (A)} Unfiltered simulation of an OU process with parameters $\lambda=1$, $\omega=60$, $\sigma=8$.
{\bf (B)} Time-frequency representation (spectrogram) using 1 second window and 75\% overlap.
{\bf (C)} Power spectrum density (blue) in the observed window $T=20s$ and the total signal window $T=1200s$ (red).
{\bf (D)} Spindles present in the $\alpha-$band.
{\bf (E)} Spectrogram computed in the $\alpha-$band.
}}
\label{fig:1}
\end{figure*}

A two-dimensional Ornstein-Uhlenbeck (OU) process is governed by the stochastic differential equation
\beq \label{ou_eq}
\dot{\s}=A\s+ \sqrt{2\sigma}\dot{\w},
\eeq
where the anti-symmetric matrix is
\[ A=
\begin{pmatrix}
-\lambda & \omega \\
-\omega &-\lambda \\
\end{pmatrix} \]
associated to the state variable $\s=(x,y) \in \rR^2$. The dynnamics is driven by a two-dimensional Brownian motion $\w$ with mean zero and variance one.  \\
The matrix $A$ has complex conjugate eigenvalues $\mu_{\pm} = -\lambda \pm i\omega$, indicating that in the absence of noise (i.e., the deterministic limit), trajectories spiral inward toward the origin. These trajectories exhibit exponentially decaying oscillations, with decay rate $\lambda$ and angular frequency $\omega$. As we shall see in Fig. \ref{fig:1},the noise term drives the system away from the fixed point, generating stochastic trajectories that repeatedly spiral around the origin, thereby producing transient oscillations with amplitude modulated by the noise intensity.\\
The transient statistical properties of the components $x(t)$ and $y(t)$ of the OU process can be explored numerically. To this end, we simulate trajectories using a fourth-order Runge-Kutta scheme (Fig.~\ref{fig:1}A), across a range of parameters that produce noise-driven bursting oscillations, as illustrated in Fig.~\ref{fig:1}A--B. Remarkably, both $x(t)$ and $y(t)$ exhibit recurrent transient bursts, which become apparent when analyzed using time-frequency representations.

Specifically, the spectrogram—computed via a sliding window Fourier transform (1-second window with 75\% overlap)—reveals intermittent oscillatory activity centered around a dominant frequency. The corresponding power spectral density (PSD) exhibits a pronounced peak at the natural frequency $\omega$ of the deterministic system (Fig.~\ref{fig:1}B--C). These observations confirm that the stochastic forcing in the OU process generates a continuous power spectrum concentrated around the resonant frequency $\omega$. This behavior is further supported by the analytical expression for the PSD, which we derive below and provide in full in Appendix~A.
\begin{figure*}
\centering
\includegraphics[width=0.9\linewidth]{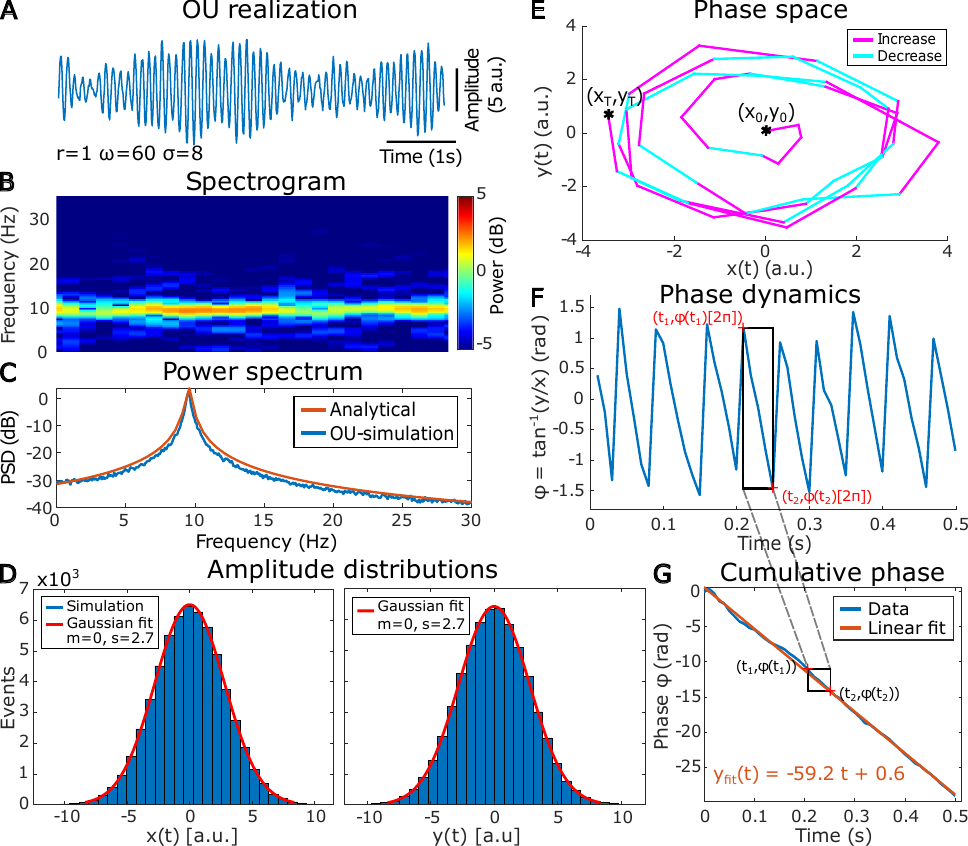}
\caption{{\bf Properties of the two dimensional Ornstein-Uhlenbeck process.}
{\small
{\bf (A)} Simulation of the Ornstein-Uhlenbeck process with parameters $\lambda=1$, $\omega=60$, $\sigma=8$.
{\bf (B)} Time-frequency representation (spectrogram) of the simulation in (A).
{\bf (C)} Power spectrum (blue) of the simulation in (A) and its theoretical fit (orange) obtained in eq.(\ref{psd}).
{\bf (D)} Distribution of the amplitude of the signal (blue) and its Gaussian fit (red) for the first and second component $x(t)$ and $y(t)$.
{\bf (E)} Phase-space between the process components for 1 second simulation. The distance from the point $(x,y)$ to the center $(0,0)$ increases (light blue) and decreases (magenta).
{\bf (F)} Evolution of the phase between $y(t)$ and $x(t)$.
{\bf (G)} Evolution of the cumulative phase between $y(t)$ and $x(t)$ (blue), fitted by a linear curve (brown).
}}
\label{fig:2}
\end{figure*}
We now turn to the analysis of the steady-state distribution of the process defined in Eq.~\eqref{ou_eq} (Fig.~\ref{fig:1}A--C), which can be derived from the stationary solution of the forward Fokker–Planck equation \cite{gardiner1985handbook,risken1996fokker,schuss2009theory}.The probability density function (pdf)
\beq \label{pxTtau}
p(\s,t\,|\,\s_0)\,d\x = \pP(\s(t)\in\x+d\x|\,\s_0).
\eeq
is solution of the Fokker-Planck equation (FPE)
\beq \label{FPEp}
\frac{\p p(\s,t\,|\,\s_0)}{\p t}=\sigma \Delta p(\s,t\,|\,\s_0) -\sum_{i=1}^d\frac{\p b^i(\s)p(\s,t\,|\,\s_0)}{\p x^i},
\eeq
where the vector field $\Bb(\s)=A\s$ has components
\beq
\mb{b}=
\begin{pmatrix}
-\lambda x+\omega y \\
-\omega x -\lambda y \\
\end{pmatrix}.
\eeq
The time-dependent solution of the Forward Fokker-Planck can be expanded on eigenfunctions, associated with the  eigenvalues $-\lambda(n+m)-i(m-n)\omega$ \cite{chen2014eigenfunctions}. The solution is given by
\beq \label{spectral}
p(\s,t\,|\,\s_0)=\sum_{n,m=0}^{\infty} c_{n,m} (\s_0) e^{-\lambda (n+m) t}(\cos((m-n)\omega t) \Re e(J_{m,n}(\s))+\sin((m-n)\omega t) \Im m(J_{m,n}(\s))),
\eeq
where $J_{m,n}$ are given by (with variable $\s=(x,y)$),
\beq
J_{m,n}(x, y)=\left\{
  \begin{array}{@{}ll@{}}
    (-1)^n n! (x+iy)^{m-n} L^{m-n}_{n} (||\s||_2^2,\rho), &\text{ if } m \geq n\\
    (-1)^m m! (x-iy)^{n-m} L^{n-m}_{m} (||\s||_2^2,\rho) &\text{otherwise.}
  \end{array}\right.
\eeq
Here the parameter $\rho=\frac{\lambda}{2\sigma}$ and the Laguerre polynomials are defined by the formula \cite{silverman1972special}
\beq
L^\alpha_n(x,\rho) =\frac{\rho^n}{n!} x^{-\alpha}  e^{x/\rho} \frac{d^n}{dx^n} e^{-x/\rho} x^{n+\alpha}, \, n \in \mathbb{N}^*.
\eeq
Unfortunately, the spectral decomposition given in Eq.~\eqref{spectral} is insufficient to capture the statistics associated with the transient, randomly occurring spindle oscillations illustrated in Fig.~\ref{fig:1}--\ref{fig:2}. This limitation suggests that the classical eigenfunction expansion, while valuable for analyzing transient from a given state or stationary behavior, does not adequately describe the non-stationary features of spindle dynamics.

To address this, the remainder of the manuscript introduces an alternative approach for quantifying spindle statistics. Specifically, we focus on characterizing each spindle in terms of its increasing and decreasing phases, which we extract using a segmentation procedure based on Empirical Mode Decomposition (EMD). This method allows us to access time-resolved properties of individual spindle events and to compute their relevant statistical distributions.
\subsection{Steady-state solution of the Fokker-Planck Equation}
The steady-state solution of the Fokker-Planck Equation (FPE) eq. \ref{FPEp}, is a Gaussian \cite{risken1996fokker}
\beq
f(\s)=\frac{1}{2\pi \sqrt{\det R}}\exp\left(-\frac{1}{2} \s^T R^{-1}\s\right),
\eeq
where the matrix $R$ satisfies the Lyapunov's equation \cite{schuss2011nonlinear}
\beq \label{Lya}
(-A)R+R (-A)^T= 2 \sigma I_{2},
\eeq
where $I_2$ is the identity matrix in dimension 2.  The matrix solution of eq. \ref{Lya} is
\beq
R=\frac{\sigma}{\lambda} I_2,
\eeq
and thus the steady-state solution is given by
\beq \label{gaussianequil}
f(\s)= \frac{\lambda}{2\pi \sigma}\exp\left(-\frac{1}{2}\frac{\lambda|\s|^2}{\sigma}\right).
\eeq
At this stage, we conclude that the $y$-projection of the OU process can be used to estimate the ratio $\frac{\sigma}{\lambda}$ by fitting a Gaussian distribution to a single stochastic realization, as illustrated in Fig.~\ref{fig:1}D (red curves). This reflects the fact that each marginal component of the OU-process reaches a stationary Gaussian distribution whose variance depends on the noise amplitude and damping coefficient.

Another informative quantity is the radial process $R(t) = \sqrt{x^2(t) + y^2(t)}$, which provides a natural way to distinguish the waxing and waning phases of spindle-like oscillations. Specifically, we classify time points as part of the increasing phase when $R(t+\Delta t) > R(t)$ (magenta) and the decreasing phase when $R(t+\Delta t) < R(t)$ (blue). Numerical simulations reveal that these phases are approximately balanced in duration, suggesting that the interplay between the deterministic attraction toward the origin and stochastic perturbations leads to a quasi-symmetric oscillatory envelope centered around the resonance frequency $\omega$.

To further investigate this behavior, we examine the distribution of angular velocities and phase trajectories in the $(x, y)$ plane. Filtering the signal around the frequency $\omega$ reveals the emergence of distinct spindle events and their confinement to a narrow frequency band (Fig.~\ref{fig:1}D--E), consistent with the underlying resonant dynamics.
\subsubsection{Power spectrum computationa of the two-dimensional OU-process}
To quantitatively characterize the resonance at frequency $\omega$, we compute the power spectral density of the first component $x(t)$ by applying the Fourier transform to its autocorrelation function. The explicit form of this autocorrelation function is derived in Appendix~A, yielding:
\beq \label{correlationexpressionmain}
C_{xx}(\tau) = \frac{\sigma}{\lambda} \exp(-\lambda|\tau|) \cos(\omega \tau).
\eeq
We recall that the power spectral density is connected to the correlation function  by \cite{schuss2009theory}
\beq \label{psd}
S(f) &=& \frac{1}{2\pi} \int_{\,\rR} C_{xx}(\tau) \exp(-i2 \pi f \tau) \, \mathrm{d}\tau \nonumber \\
     &=& \frac{\sigma}{4\pi \lambda}  \, \int_{\,\rR} (\exp(i\omega \tau) + \exp(-i\omega \tau)) \exp(-\lambda|\tau| -i2\pi f\tau) \, \mathrm{d}\tau.
\eeq
A direct integration of eq.(\ref{psd}) leads to
\beq
S(f) &=& S_1(f) + S_2(f),
\eeq
where
\beq \label{psd_1}
S_1(f) &=& \frac{\sigma}{4\pi \lambda} \,\int_{\,\rR^{-}} (\exp((\lambda+i(\omega-2\pi f))\tau) + \exp((\lambda-i(\omega+2\pi f))\tau)) \, \mathrm{d}\tau \nonumber \\
       &=& \frac{\sigma}{4\pi \lambda} \, \left(\frac{1}{\lambda+i(\omega-2\pi f)} + \frac{1}{\lambda-i(\omega+2\pi f)} \right) +CC,
\eeq
where CC means complex conjugated.
Similarly,
\beq \label{psd_2}
S_2(f) &=& \frac{\sigma}{4\pi \lambda} \int_{\,\rR^{+}} (\exp((-\lambda+i(\omega-2\pi f))\tau) + \exp((-\lambda-i(\omega+2\pi f))\tau)) \, \mathrm{d}\tau \nonumber \\
       &=& \frac{\sigma}{4\pi \lambda} \, \left(\frac{1}{\lambda-i(\omega-2\pi f)} + \frac{1}{\lambda+i(\omega+2\pi f)} \right)
\eeq
Finally, we obtain the simplified expressions (\ref{psd_1}) and (\ref{psd_2})
\beq \label{pspectrum}
S(f) 
     = \frac{\sigma}{2\pi} \left(\frac{1}{\lambda^2 + (\omega-2\pi f)^2} + \frac{1}{\lambda^2 + (\omega+2\pi f)^2}\right)
\eeq
that we used to fit the spectral density (Fig. \ref{fig:2}C). The maximum resonance occurs at frequency $f_{res} = \frac{\omega}{2\pi}$. Finally, the total energy is given by  (\ref{pspectrum})
\beq
E_{tot} = \int_{\, \rR} S(f) \,\mathrm{d}f
        = \frac{\sigma}{2 \pi \lambda}.
\eeq
\subsubsection{Width of the Power Spectral Density at half maximum power}
To quantify the frequency dispersion induced by the spindle dynamics, we use the full width at half maximum (FWHM) of the power spectral density. The FWHM is defined as the bandwidth over which the power remains above half of its peak value, providing a measure of the concentration of energy around the dominant frequency. For the power spectrum given in Eq.~\eqref{pspectrum}, the maximum occurs at the resonant frequency $f_{\mathrm{max}} = \frac{\omega}{2\pi}$, and the corresponding peak power is
\beq
S(f_{max}) = \frac{\sigma}{2\pi} \left(\frac{1}{\lambda^2} + \frac{1}{\lambda^2+4\omega^2} \right).
\eeq
The width is given by the solution of equation
\beq
S(f) = \frac{1}{2} S(f_{max}).
\eeq
The two solutions are given by
\beq
f_{\pm} = \frac{1}{2\pi}\sqrt{\frac{a(\lambda,\omega) \pm b(\lambda,\omega)}{c(\lambda,\omega)}}
\eeq
where
\beq
a(\lambda,\omega) &=& 3\lambda^2 \omega^2 + 2\omega^4, \nonumber \\
b(\lambda,\omega) &=& \lambda\sqrt{\lambda^6+8\lambda^4\omega^2+24\lambda^2\omega^4+16\omega^6} \\
c(\lambda,\omega) &=& \lambda^2+2\omega^2. \nonumber
\eeq
Finally, the width of the $\alpha-$band is given by
\beq
\FWHM = \frac{1}{2\pi} \left(\sqrt{\frac{a(\lambda,\omega)+b(\lambda,\omega)}{c(\lambda,\omega)}} - \sqrt{\frac{a(\lambda,\omega)-b(\lambda,\omega)}{c(\lambda,\omega)}}\right).
\eeq
When $\omega \gg \lambda$, we obtain the following asymptotic expression
\beq
\FWHM \sim  \frac{\lambda}{\pi}+\frac{\lambda^3}{4\pi \omega^2}+\frac{11 \lambda^5}{32 \pi \omega^2}+O\left(\lambda^6\right)
\eeq
In summary, the width of the resonance band is primarily governed by the damping parameter $\lambda$, which controls the rate of decay toward the fixed point. Our numerical simulations confirm that a two-dimensional OU process, under appropriate noise levels, can spontaneously generate spindle-like oscillations with substantial amplitude and a well-defined frequency structure. These results support the use of this minimal stochastic model for capturing the essential features of random burst dynamics observed in empirical signals.
\subsection{Range of parameters leading to spindles vs oscillations}
Spontaneous transient spindles emerge within a specific range of the Ornstein-Uhlenbeck (OU) process parameters $(\lambda, \omega, \sigma)$ as defined in Eq.~\eqref{ou_eq}. To generate pronounced bursting behavior, certain conditions must be met. In particular, one must avoid regimes where the dynamics are dominated by small fluctuations around the origin, as occurs when the damping parameter $\lambda$ is too small \cite{thomas2019phase,verechtchaguina2007interspike}. Instead, sustained spindle-like oscillations require an appropriate balance between rotational drift and stochastic excitation. This leads to the following criteria:
\begin{enumerate}
  \item The eigenvalues of the matrix $A$ in Eq.~\eqref{ou_eq} must be complex, i.e., $\omega > 0$, ensuring rotational dynamics around the fixed point.
  \item The system should exhibit underdamped behavior, meaning the oscillatory timescale dominates over the relaxation time: $\omega / \lambda \gg 1$.
  \item The noise amplitude $\sigma$ must be sufficiently large to induce excursions away from the origin, but not so large as to destroy the coherent bursting structure.
\end{enumerate}
Under these conditions, the process frequently escapes the vicinity of the fixed point, and the empirical distribution over any time window of duration greater than $1/\omega$ deviates significantly from the equilibrium density centered at $(x, y) = (0, 0)$.
\section{Spindles decomposition and analysis on OU process}
Stochastic simulations reveal that spindle results from the noise that amplify the converging deterministic oscillation and pushes trajectories to an almost periodic motion in the opposite direction of the attractor (purple line in Fig. \ref{fig:2}E, as opposed to the cyan lines). We will use the Gaussian solution in eq.(\ref{gaussianequil}) below to estimate the relevant parameters, but it is generally a poor description of the spindle dynamics generated by the OU dynamics. We also fitted the real and imaginary part of the Hilbert transformed of the signal $x(t)$. We recall that Hilbert-Huang \cite{flandrin2004empirical,huang1998empirical} transform is defined by
\beq
\mathcal{H}(x)(t) = \frac{1}{\pi},\hbox{p.v.} \int_{-\infty}^{\infty} \frac{x(\tau)}{t-\tau} \mathrm{d}\tau,
\eeq
where p.v. is the Cauchy principal value. We computed $y(t)=\mathcal{H}\{x\}(t)$ and obtain the analytical signal $x_a(t)$ defined by
\beq \label{EMDdecomp}
x_a(t) &=& x(t) + i y(t) \nonumber \\
       &=& A(t) \exp(i\phi(t)),
\eeq
where $\phi(t)$ is the phase and $A(t)$ the instantaneous amplitude.
We first computed the phase of the OU process (\ref{ou_eq}), defined as $\phi(t)=\arctan\left(\frac{y(t)}{x(t)}\right)$. As shown in Fig. \ref{fig:2}F, the phase decreases proportionally with $\omega$. We note the resetting each time the variable $x$ crosses y-axis). The linear decay can be found by applying Ito's formula to $f(x(t),y(t))= \arctan\left(\frac{y(t)}{x(t)}\right)$:
\beq
d\phi = -\omega \mathrm{d}t +\frac{\sqrt{2\sigma}}{A(t)}\mathrm{d}\tilde w,
\eeq
where $r=dist (\s,0)$ and $\tilde w$ is a Brownian motion.
We show that deterministic component has a linear decay with slope $-\omega$ (Fig. \ref{fig:2}G)
Interestingly, when the distance $A$ diverges from zero, the oscillation is driven by the deterministic drift $\omega$. When the distance $A$ goes near zero, the Brownian term dominates and dictate the envelope behavior.
\subsection{Spindle segmentation} \label{s:segmentation}
\begin{figure*}[http!]
\vspace{-2cm}
\centering
\includegraphics[width=0.75\linewidth]{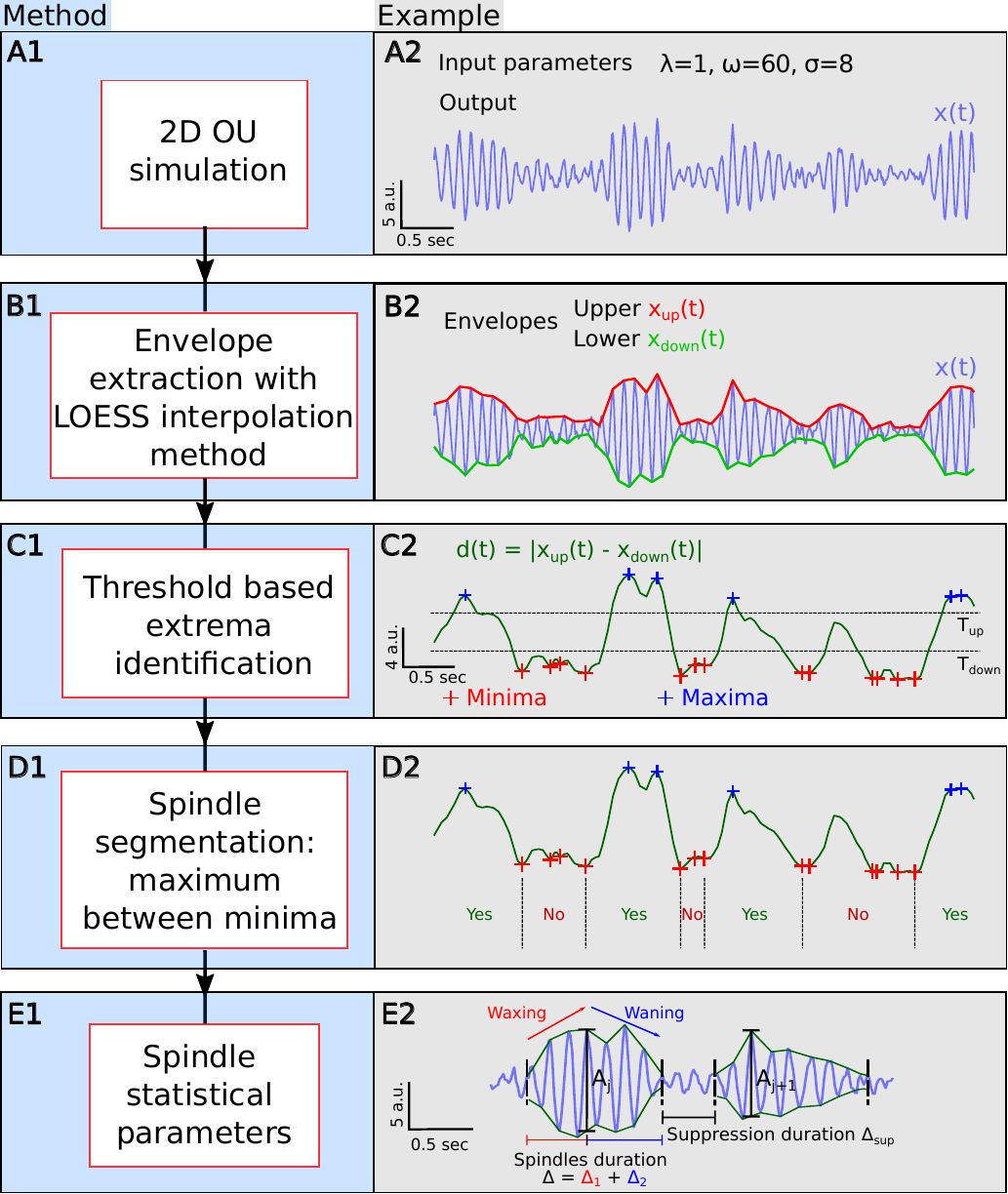}
\caption{{\bf Spindles segmentation algorithm}
{\small
{\bf (A)} Signal $x(t)$ and its filtered signal $x_\alpha(t)$ in the $\alpha-$band frequency range (8-12 Hz).
{\bf (B)} Envelope extraction using locally weighted linear regression with second-degree polynomial model (LOESS) on the signal local minima (green) and maxima (red)
{\bf (C)} Envelopes distance (dark green) and their local minima (red) (resp. maxima (blue)) under (resp. above) a given threshold $T_{up}$ (resp. $T_{down}$).
{\bf (D)} Spindle detection criteria based on the presence of at least one local maxima between two successive minima.
{\bf (E)} Statistical properties calculation: spindle increase $\Delta_1$ and decrease $\Delta_2$ phases duration, total spindle duration $\Delta$ and inter-spindle duration $\Delta_{sup}$.
}}
\label{fig:3}
\end{figure*}

In the absence of a universally accepted definition of a spindle, we propose to define it as a transient oscillatory event occurring within a continuous signal, delimited by two successive local minima in its amplitude envelope. To formalize this, we express the signal $x(t)$ in complex coordinates as
\beq
x(t) = A(t) \exp(i\phi(t)),
\eeq
where $A(t)$ denotes the slowly varying amplitude (envelope), and $\phi(t)$ encodes the fast oscillatory phase component. The scale separation assumption $\frac{d}{dt}A(t) \ll \frac{d}{dt}\phi(t) \approx \omega$ ensures that the amplitude captures low-frequency modulation, while the phase evolves on a faster timescale set by the dominant frequency $\omega$.\\
In practice, we shall estimate the envelope $A(t)$ by applying Empirical Mode Decomposition (EMD) to the signal $x(t)$ \cite{huang1998empirical,flandrin2004empirical}. Based on this representation, we define a spindle as the segment of the signal between two successive local minima of $A(t)$ that fall below a threshold $T_{\text{down}}$, and that enclose a single maximum exceeding a higher threshold $T_{\text{up}}$ (Fig.~\ref{fig:3}). The spindle segmentation proceeds via the following steps:
\begin{enumerate}
    \item \textbf{Identification of local extrema.}
    Detect all local maxima and minima of the signal $x(t)$ using Empirical Mode Decomposition (EMD). The envelope of the signal is extracted via locally weighted polynomial regression (LOESS) using a second-degree model \cite{jacoby2000loess}.

    \item \textbf{Envelope smoothing to reduce sampling noise.}
    Construct the upper envelope $x_{\text{up}}(t)$ and the lower envelope $x_{\text{down}}(t)$ by applying LOESS smoothing to the sequences of local maxima and minima, respectively.

    \item \textbf{Computation of the envelope amplitude.}
    Define the instantaneous envelope amplitude $d(t)$ as the difference between the smoothed upper and lower envelopes:
    \begin{equation}
    d(t) = |x_{\text{up}}(t) - x_{\text{down}}(t)|.
    \end{equation}

    \item \textbf{Threshold-based segmentation of extrema.}
    Identify time points where the amplitude envelope reaches prominent peaks and troughs by thresholding:
    \begin{itemize}
        \item Define a minimum threshold $T_{\min} = \sigma_x$ and a maximum threshold $T_{\max} = 3\sigma_x$, where $\sigma_x$ is the standard deviation of the signal $x(t)$.
        \item Identify the set of local maxima of $d(t)$ that exceed $T_{\max}$:
        \begin{equation}
        \mathcal{S}_{T_{\max}} = \left\{ t_i \;\middle|\; \exists \varepsilon > 0,\, \forall t \in [t_i - \varepsilon, t_i + \varepsilon],\, d(t_i) > d(t),\, d(t_i) > T_{\max} \right\}.
        \end{equation}
        \item Similarly, identify the set of local minima of $d(t)$ that fall below $T_{\min}$:
        \begin{equation}
        \mathcal{S}_{T_{\min}} = \left\{ t_j \;\middle|\; \exists \varepsilon > 0,\, \forall t \in [t_j - \varepsilon, t_j + \varepsilon],\, d(t_j) < d(t),\, d(t_j) < T_{\min} \right\}.
        \end{equation}
    \end{itemize}

    \item \textbf{Spindle definition.}
    A spindle is defined as the segment of the signal between two consecutive time points in $\mathcal{S}_{T_{\min}}$ that contains at least one local maximum in $\mathcal{S}_{T_{\max}}$. Formally,
    \begin{equation} \label{spindledef}
    \mathcal{S}_{\text{Spindles}} = \bigcup_{(t_i, t_{i+1}) \in \mathcal{S}_{T_{\min}}} \left\{ x(t) \;\middle|\; t \in [t_i, t_{i+1}],\, \exists t_j \in \mathcal{S}_{T_{\max}},\; t_i < t_j < t_{i+1} \right\}.
    \end{equation}
\end{enumerate}
The procedure is illustrated in Fig. \ref{fig:4}.
This segmentation procedure can be formalized in the following pseudo-code, which identifies spindle intervals based on thresholded extrema of the amplitude envelope. The thresholds are estimated using LOESS interpolation, and the detection can be adapted to real-time by applying the algorithm within a sliding window $[t, t+T]$ (see remark below).
\begin{algorithm}[H]
\caption{Spindle Segmentation Algorithm (SSA)} \label{alg:SSA}
\textbf{Input:} Time series signal $x(t)$ (e.g., EEG) \\
\textbf{Output:} \texttt{Spindles} — list of detected spindle intervals
\begin{algorithmic}[1]
\State Initialize \texttt{Spindles} $\gets [\,]$
\State Identify all local maxima and minima of $x(t)$
\State Apply LOESS smoothing to local maxima and minima to obtain $x_{\text{up}}(t)$ and $x_{\text{down}}(t)$
\State Compute dynamic range: $d(t) = |x_{\text{up}}(t) - x_{\text{down}}(t)|$
\State Set amplitude thresholds: $T_{\text{down}} = \operatorname{std}(d)$, $T_{\text{up}} = 3 \cdot \operatorname{std}(d)$
\State Extract local minima $\texttt{LMIN} = \{t_{\min} \mid d(t_{\min}) < T_{\text{down}}\}$
\State Extract local maxima $\texttt{LMAX} = \{t_{\max} \mid d(t_{\max}) > T_{\text{up}}\}$
\For{$i = 1$ to $\text{len}(\texttt{LMIN}) - 1$}
    \State $t_{\text{start}} \gets \texttt{LMIN}[i]$, $t_{\text{end}} \gets \texttt{LMIN}[i+1]$
    \If{there exists $t_{\max} \in \texttt{LMAX}$ such that $t_{\text{start}} < t_{\max} < t_{\text{end}}$}
        \State Append $[t_{\text{start}}, t_{\text{end}}]$ to \texttt{Spindles}
    \EndIf
\EndFor
\State \Return \texttt{Spindles}
\end{algorithmic}
\end{algorithm}
\noindent\textbf{Remark.} For real-time or streaming applications, the algorithm can be applied within a sliding window $[t, t+T]$ over the full signal duration. Thresholds and LOESS fits should be updated locally within each window to adapt to potential nonstationarities in the amplitude envelope.
\subsection{Robustness of spindle segmentation algorithm}
\begin{figure*}[http!]
\centering
\includegraphics[width=1\linewidth]{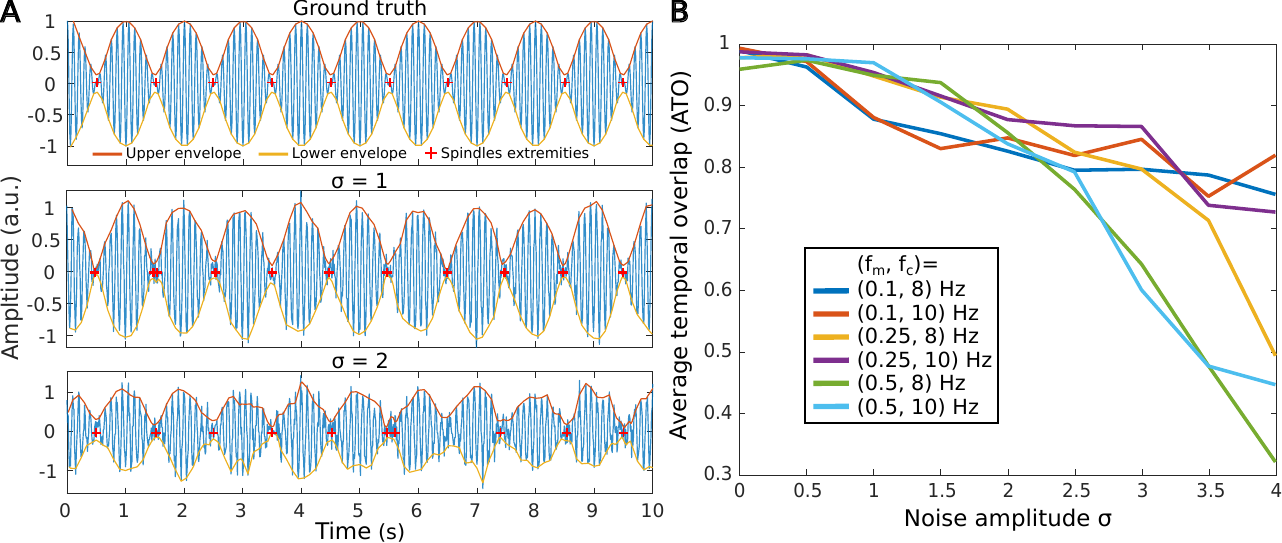}
\caption{{\bf Spindles segmentation performance based on SSA.}
{\small
{\bf (A)} Spindles segmentation examples for various noise amplitude $\sigma=$ 0,1,2.
{\bf (B)} Curves of the average temporal overlap (ATO) between the segmented spindles and the ground truth spindles versus the noise amplitude $\sigma$ for different modulating (0.1, 0.25 and 0.5 Hz) and carrier (8 and 10 Hz) frequencies.
}}
\label{fig:4}
\end{figure*}

To evaluate the robustness of the spindle segmentation algorithm (SSA) introduced above, we tested it on a set of ground truth analytical spindle defined by
\beq
Sp(t) = \sin(2\pi f_m t)\cos(2\pi f_c t)
\eeq
where the frequency $f_m$ is a modulating frequency and $f_c$ is the carrier frequency. To simulate a proper spindle (Fig. \ref{fig:4}A), we choose two frequencies with a gap $f_m=0.5$ Hz and $f_c=10$ Hz.
The ground truth spindle intervals are given by
\beq
\mathcal{S}_{GT} = \bigcup_{k \; \in \; \nN} \left\{ Sp(t) \; \hbox{ for } t \in \left[\frac{k}{2 f_m},\frac{k+1}{2 f_m}\right] \right\}.
\eeq
We then added a centered Brownian noise $w$ with standard deviation $\sigma$ to the deterministic spindles so that the perturbed signal is
\beq \label{testfunction}
S_{random}(t) = \sin(2\pi f_m)\cos(2\pi f_c t) + \sigma w(t),
\eeq
leading to perturbed spindles (Fig. \ref{fig:4}A). To test the performance of the SSA, we compared the segmented spindles and quantified the difference between perturbed and unperturbed spindles by computing the average temporal overlap (ATO) defined as
\beq
ATO =\frac{1}{N} \sum_{i=1}^{N} \frac{P_i \cap GT_i}{P_i \cup GT_i}
    = \frac{1}{N} \sum_{i=1}^{N} \frac{\min(t_{P_i,end},t_{GT_i,end}) - \max(t_{P_i,start},t_{GT_i,start})}{\max(t_{P_i,end},t_{GT_i,end}) - \min(t_{P_i,start},t_{GT_i,start})},
\eeq
which is the average of the ratio of the time segments intersection and its union with $N = Card(\mathcal{S}_{GT})$.
When we varied the added Brownian noise amplitude $\sigma$ between 0 to 4 times the signal amplitude, we obtained that the algorithm was able under low noise amplitudes ($\sigma < 2$), to recover the spindle segmentation. However, its efficacy decreases with higher noise amplitudes $\sigma > 2$ (Fig. \ref{fig:4}B). Finally, these results remain consistent across different choices of modulating and carrier frequencies. To conclude, the present segmentation algorithm is robust to noise amplitude.
\section{Stochastic spindle statistics}
In this section, we analyze the stochastic dynamics underlying spindle amplitudes generated by the two-dimensional Ornstein--Uhlenbeck (OU) process \eqref{ou_eq}. We derive asymptotic and explicit expressions for the distributions of spindle duration, amplitude, and the increasing and decreasing  phases (Fig.~\ref{fig:3}E). Spindles are understood here as stochastic excursions in amplitude, modulated by the noisy dynamics of the OU system.

Starting from  the components $(x(t), y(t))$ of the OU process, we introduce polar coordinates:
\beq
x(t) + i y(t) = r(t) e^{i \phi(t)},
\eeq
where $r(t)$ is the instantaneous amplitude and $\phi(t)$ is the phase. Applying Ito's formula to the SDEs satisfied by $x(t)$ and $y(t)$ yields the following stochastic dynamics:
\begin{align}
\mathrm{d}x^2 &= [2x(-\lambda x + \omega y) + 2\sigma] \mathrm{d}t + 2x\sqrt{2\sigma} , \mathrm{d}w_1, \
\mathrm{d}y^2 &= [2y(-\lambda y - \omega x) + 2\sigma] \mathrm{d}t + 2y\sqrt{2\sigma} , \mathrm{d}w_2,
\end{align}
where $w_1$ and $w_2$ are independent Brownian motions.
Summing the two equations gives the evolution of $r^2(t)$:
\begin{equation}
\mathrm{d}r^2 = (-2\lambda r^2 + 4\sigma) , \mathrm{d}t + 2r\sqrt{2\sigma} , \mathrm{d}\omega,
\end{equation}
where $\omega(t)$ is a standard Brownian motion. Setting $u = r^2$, we obtain:
\begin{equation} \label{stochasticdist2}
\mathrm{d}u = (-2\lambda u + 4\sigma) , \mathrm{d}t + 2\sqrt{2\sigma u} , \mathrm{d}w.
\end{equation}
Applying Ito's formula to $f(u) = \sqrt{u} = r$ yields the SDE for $r(t)$:
\begin{equation} \label{stochasticdist3}
\mathrm{d}r = \left(-\lambda r + \frac{\sigma}{r}\right) \mathrm{d}t + \sqrt{2\sigma} , \mathrm{d}w.
\end{equation}
This equation can be written in the form of a gradient flow:
\begin{equation} \label{stochasticdist}
\mathrm{d}r = -W'(r) , \mathrm{d}t + \sqrt{2\sigma} , \mathrm{d}w,
\end{equation}
with effective potential:
\begin{equation}
W(r) = \frac{\lambda}{2} r^2 - \sigma \log r.
\end{equation}
The equilibrium amplitude $r_{\mathrm{eq}}$ satisfies $W'(r_{\mathrm{eq}}) = 0$, giving:
\begin{equation}
r_{\mathrm{eq}} = \sqrt{\frac{\sigma}{\lambda}}.
\end{equation}
The corresponding Fokker--Planck equation for the probability density $p(r, t)$ reads:
\begin{align} \label{FPE}
\frac{\partial p}{\partial t}(r, t) &= \sigma \frac{\partial^2 p}{\partial r^2}(r, t) + \frac{\partial}{\partial r}[W'(r) p(r, t)], \\
p(r, 0) &= \delta(r),
\end{align}
where $\delta$ is the Dirac delta function. The stationary solution satisfies:
\begin{equation}
p(r) = C \exp\left(-\frac{W(r)}{\sigma}\right),
\end{equation}
with normalization constant:
\begin{align}
C &= \frac{1}{\int_0^\infty x \exp\left(-\frac{\lambda}{2\sigma} x^2\right) \mathrm{d}x} = \frac{\lambda}{\sigma}.
\end{align}
Hence, the stationary probability density function of the spindle amplitude $r$ is:
\begin{equation} \label{pdfSteadtstate}
p(r) = \frac{\lambda}{\sigma} r \exp\left(-\frac{\lambda}{2\sigma} r^2\right).
\end{equation}

We now compute the expectation and variance of $r$. Using the standard integrals:
\begin{align}
\int_0^\infty r^2 e^{-b r^2} \mathrm{d}r = \frac{1}{4b} \sqrt{\frac{\pi}{b}}, \
\int_0^\infty r^3 e^{-b r^2}  \mathrm{d}r = \frac{1}{2b^2},
\end{align}
and setting $b = \frac{\lambda}{2\sigma}$, we find:
\begin{align}
\mathbb{E}[r] = \sqrt{\frac{\pi \sigma}{2\lambda}}, \
\mathrm{Var}(r) = 2 \frac{\sigma}{\lambda} - \frac{\pi \sigma}{2\lambda} = \frac{\sigma}{\lambda} \left(2 - \frac{\pi}{2}\right).
\end{align}
These expressions provide explicit insights into the amplitude distribution of stochastically generated spindles. In particular, they show that the mean amplitude scales as $\sqrt{\sigma / \lambda}$, while the variance is controlled by a balance between noise intensity and damping.
\subsection{Estimation of inter-spindle MFPT duration}\label{interspindleestimation}
To estimate the mean inter-spindle duration, we compute the \textit{mean first passage time} (MFPT) for the amplitude process $r(t)$ to reach a threshold $T > 0$, which marks the onset of a spindle. The expected time $u(y)$ for the process starting at $r(0) = y$ to reach $T$ satisfies the following second-order differential equation (Dynkin's formula) \cite{schuss2009theory}:
\begin{align} \label{MFPT}
\sigma \frac{d^2 u}{dy^2}(y) - W'(y) \frac{d u}{dy}(y) &= -1, \quad 0 < y < T, \\
u(T) &= 0, \quad \text{(absorbing)}, \nonumber \\
u'(0) &= 0, \quad \text{(reflecting)}. \nonumber
\end{align}
Here, $W(y) = \frac{\lambda}{2} y^2 - \sigma \log y$ is the effective potential for the radial amplitude process as derived previously in Eq.~\eqref{stochasticdist}. The boundary condition at $y = 0$ models a reflecting origin, consistent with the fact that $r(t) \geq 0$ and the process cannot cross zero.\\
A first integration of Eq.~\eqref{MFPT} yields
$ u'(y) = \frac{1}{\lambda y} \left( 1 - \exp\left( \frac{\lambda y^2}{2\sigma} \right) \right).$
Integrating again from a starting point $a$ to threshold $T$ gives:
\begin{equation}
u(a) = \int_a^T \frac{1}{\lambda y} \left( \exp\left( \frac{\lambda y^2}{2\sigma} \right) - 1 \right) dy.
\end{equation}
In particular, when starting from $a = 0$, we perform the change of variable $r = \sqrt{\frac{\lambda}{2\sigma}} y$ to obtain:
\begin{equation} \label{MFPTactivationSpind}
u(0) = \frac{1}{\lambda} \int_0^{\sqrt{\frac{\lambda}{2\sigma}} T} \frac{1}{r} \left( \exp(r^2) - 1 \right) dr.
\end{equation}
To analyze the behavior for small thresholds, we consider $T = \sqrt{g \sigma}$, where $g = \mathcal{O}(1)$. Expanding the exponential for small $r$ gives:
\begin{align}
u(0) \approx \frac{1}{\lambda} \int_0^{\sqrt{g/2}} \frac{1}{r} \left(1 + r^2 + \frac{r^4}{2} + \cdots - 1 \right) dr \approx \frac{1}{\lambda} \int_0^{\sqrt{g/2}} r \, dr = \frac{1}{2\lambda} \cdot \frac{g}{2} = \frac{g}{4\lambda}.
\end{align}
Thus, for $g = 2$ and $\lambda = 1$, the MFPT is approximately $u(0) \approx 1/2 s$. Due to the symmetry of the dynamics, the total expected time between the end of one spindle and the initiation of the next includes both the activation and return times. Hence, the mean inter-spindle interval is of order one, independent of the noise amplitude $\sigma$:
\begin{equation}
\bar{\tau}_{\text{inter-spindle}} \sim \mathcal{O}(1).
\end{equation}
This analysis highlights that spindle generation, modeled as a threshold-crossing event by a noisy radial process, occurs on a timescale set by the system’s deterministic dynamics (e.g., $\lambda$) rather than the noise amplitude itself. The exponential form of the potential prevents an arbitrarily long waiting time in low-noise regimes, contrasting with classical barrier crossing models where the mean exit time diverges exponentially as $\sigma \to 0$.
\subsection{Mean First Passage Time for Spindle Waxing and Waning}
To estimate the durations of the waxing (increasing) and waning (decreasing) phases of a spindle, we analyze the mean first passage time (MFPT) between amplitude thresholds. We assume that spindles initiate when the amplitude $r(t)$ crosses a lower threshold $T_{\text{Down}} = \sqrt{2\sigma}$. As the maximum amplitude reached during a spindle is stochastic, we define the upper threshold $T_{\text{Up}}$ using a quantile of the steady-state amplitude distribution:
\begin{equation} \label{Tupdefin}
\int_0^{T_{\text{Up}}} p(r) , dr = 1 - \alpha,
\end{equation}
where $\alpha \ll 1$. Using the distribution \eqref{pdfSteadtstate}, this yields:
\begin{equation}
T_{\text{Up}} = \sqrt{\frac{2\sigma}{\lambda} \log\left(\frac{1}{\alpha}\right)}.
\end{equation}
In practice, a typical value is $\alpha = 0.2$.

\paragraph{Waxing Phase.} To compute the mean duration of the waxing phase, we evaluate the MFPT from $T_{\text{Down}}$ to $T_{\text{Up}}$ using the MFPT equation \eqref{MFPT} with boundary conditions:
\begin{equation}
u'(T_{\text{Down}}) = 0, \quad u(T_{\text{Up}}) = 0.
\end{equation}
Integrating the corresponding differential equation yields:
\begin{equation}
u_1'(y) = \frac{1}{\lambda y} \left(1 - \exp\left(\frac{\lambda y^2}{2\sigma} - \lambda\right) \right).
\end{equation}
A second integration gives:
\begin{align}
u_1(T_{\text{Down}}) &= \int_{T_{\text{Down}}}^{T_{\text{Up}}} \frac{1}{\lambda y} \left(\exp\left(\frac{\lambda y^2}{2 \sigma} - \lambda\right) - 1 \right) dy \
&= \frac{e^{-\lambda}}{\lambda} \int_{T_{\text{Down}}}^{T_{\text{Up}}} \frac{\exp\left(\frac{\lambda y^2}{2\sigma}\right)}{y} dy - \frac{1}{\lambda} \int_{T_{\text{Down}}}^{T_{\text{Up}}} \frac{1}{y} dy.\nonumber
\end{align}
Changing variables and evaluating asymptotically for $\alpha \ll 1$, we obtain:
\begin{equation}
u_1(T_{\text{Down}}) \approx \frac{1}{2\lambda \alpha \log(1/\alpha)} e^{-\lambda} - \frac{1}{2\lambda} \log\left(\frac{1}{\lambda} \log\left(\frac{1}{\alpha}\right)\right).
\end{equation}

\paragraph{Waning Phase.} For the decreasing phase, we compute the MFPT from $T_{\text{Up}}$ down to $T_{\text{Down}}$, solving the same MFPT equation \eqref{MFPT} with boundary conditions:
\begin{equation}
u(T_{\text{Down}}) = 0, \quad u'(T_{\text{Up}}) = 0.
\end{equation}
The first integration yields:
\begin{equation}
u_2'(y) = \frac{1}{\lambda y} \left(1 - \alpha \exp\left(\frac{\lambda y^2}{2\sigma}\right)\right).
\end{equation}
Integrating this expression:
\begin{align}
u_2(T_{\text{Up}}) &= \int_{T_{\text{Down}}}^{T_{\text{Up}}} \frac{1}{\lambda y} \left(1 - \alpha \exp\left(\frac{\lambda y^2}{2\sigma}\right)\right) dy \
&= \frac{1}{\lambda} \int_{T_{\text{Down}}}^{T_{\text{Up}}} \frac{1}{y} dy - \frac{\alpha}{\lambda} \int_{T_{\text{Down}}}^{T_{\text{Up}}} \frac{\exp\left(\frac{\lambda y^2}{2\sigma}\right)}{y} dy.\nonumber
\end{align}
For large $T_{\text{Up}}$ (small $\alpha$), the second term is negligible, yielding:
\begin{equation}
u_2(T_{\text{Up}}) \approx \frac{1}{2\lambda} \log\left(\frac{1}{\lambda} \log\left(\frac{1}{\alpha}\right)\right) - \frac{1}{2 \lambda \log(\frac{1}{\alpha})} .
\end{equation}

\paragraph{Total Spindle Duration.} The expected total duration of a spindle is given by the sum:
\begin{align}
\bar{\tau}_{\text{Spindle}} &= u_1(T_{\text{Down}}) + u_2(T_{\text{Up}}) \approx \frac{1}{2\lambda \alpha \log(1/\alpha)} e^{-\lambda} - \frac{1}{2\lambda \log(1/\alpha)}.
\end{align}
These expressions quantify the inherent asymmetry in waxing and waning phases due to the nonlinearity of the potential and the shape of the amplitude distribution. As expected, higher spindle thresholds (i.e., smaller $\alpha$) lead to longer waxing durations, while the waning duration grows logarithmically in $\alpha^{-1}$.
\subsection{Time distribution of spindle increasing and decreasing phases}
We shall now compute the distribution of time spent in increasing $\uparrow$ (from minimum x to maximum y) and decreasing $\downarrow$ (from y to x) phases by solving the eigenvalue problem using the backward Fokker-Planck equation \cite{linetsky2004computing,davydov2003pricing,linetsky2004spectral,linetsky2004lookback}.  of the Cox-Ingersoll Ross (CIR) process, obtained from the stochastic equation of the envelope eq.(\ref{stochasticdist2}): we have
\beq
\dot{r}=\kappa(\theta-r)+\sqrt{\epsilon r}\dot{w},
\eeq
where
\beq
\kappa=2\lambda, \, \theta =\frac{2 \sigma}{\lambda},\, \epsilon
=8 \sigma.
\eeq
The corresponding backward Fokker-Planck equation is
\beq \label{FPE1}
\frac{\p}{\p t} p(r,t) &=& \frac{\epsilon r}{2} \frac{\p^2}{\p r^2} p(r,t) - \kappa(\theta-r)\frac{\p }{\p r}p(r,t) \\
p(r,0) &=& f(x) \nonumber,
\eeq
where $f(x)$ is an initial condition. The method of separation of variables leads to
\beq
p(x,t \mid y) = \sum_{k=0}^{\infty} c_k(y) \Phi_k(x) \exp (-\lambda_k t),
\eeq
where $\Phi_k$ are eigenfunctions of the Sturm-Liouville problem
\beq \label{eigenfunction}
\frac{\epsilon r}{2} \Phi_k''(r)+\kappa(\theta-r)\Phi_k'(r)+\lambda_k \Phi_k(r)=0.
\eeq
The solutions of eq. \ref{eigenfunction} have been studied in \cite{linetsky2004computing} and are linear combination of the Kummer ${}_1F_1(a;b;x)$ and Tricomi $U(a;b;x)$ confluent hypergeometric functions.
As we shall see below, the spectrum $\{\lambda_n\}$ of eq. \ref{eigenfunction} is  obtained by computing the roots of either the  Kummer or the Tricomi functions. The present problem can be transform into a classical spectral self-adjoint semigroups in Hilbert space with weight $m(x)=\frac{2}{\epsilon}\exp(-x)$ ensuring that the spectrum is discrete for bi-orthogonal with respect to $m$ solution of
\beq
(m(x)\Phi_n')'+\lambda_n \Phi_n=0,
\eeq
with appropriate boundary condition (see below).  To obtain the spindle time distribution, we shall consider two cases: the increasing (resp. decreasing ) from $T_{Down}$ to $T_{Up}$ (resp. $T_{Up}$ to $T_{Down}$) phases. We recall that the solution of eq. \ref{FPE1} can be expanded as
\beq
p_t(y|x)=\sum_{0}^{\infty} c_n(y) \Phi_n(x) \exp (-\lambda_n t).
\eeq
The survival pdf
\beq
S(t|x)= \pP(T_{x}>t)=\int_{e1}^{e2} p_t(y|x)dy =\sum_{0}^{\infty} c_n \Phi_n(x) \exp (-\lambda_n t).
\eeq
We can obtain an expansion for the time distribution of the different phase of the spindle as the formal sum
\beq
f(t|x) =-\frac{d}{dt}S(t|x)=\sum_{k=1}^{\infty} \tilde{c}_n \lambda_n \Phi_n(x)\exp(-\lambda_n t),
\eeq
where the coefficients $ \tilde{c}_n $ will be explicitly defined in the next subsections.
Note that the orthogonality of the eigenfunctions are given with respect to the weight function $m(x)=\frac{2}{\epsilon}\exp(-x)$ on interval $[0,y]$ (where $y>0$). We will compute below the spectrum using the CIR processes \cite{linetsky2004spectral}.
\subsection{Time distribution of spindle increasing phase}\label{s:waxing}
We now derive here explicit asymptotic expressions for the durations of the increasing and decreasing phases. The CIR process in eq.\ref{stochasticdist2} is defined for the square distance $r^2(t)$. Thus we shall use as boundary condition the square threshold $u=T_{Down}^2$ and $u=T_{Up}^2$ in the boundary conditions. We shall use the normalized variables
\beq
a = -\frac{\lambda}{\kappa}, \; b= \frac{2\kappa\theta}{\epsilon}=1, \; \bar{x}=\frac{2\kappa T_{Down}^2}{\epsilon}, \; \bar{y}=\frac{2\kappa T_{Up}^2}{\epsilon}
\eeq
so that the solution of \ref{eigenfunction}
\beq
z v''+(b-z)v'-av=0.
\eeq
is given by
\beq
v(z)=A _1 F_1(a;b;z)+ B U(a;b;z),
\eeq
where A and B are two constants that are determined with the boundary conditions $v(\bar{y})=0$ and $v'(\bar{x})=0$.  Following the method developped in \cite{linetsky2004computing}, we obtain that the non-trivial solution of eq.(\ref{eigenfunction}) is the Kummer function ${}_1 F_1(a;b;\bar{x})$. The spectrum $\{\lambda_n\}$ is given by the roots of equation
\beq \label{spectrum}
{}_1 F_1(a;b;\bar{y})=0.
\eeq
The coefficients become
\beq
c_n = -\frac{{}_1 F_1(a_n;b;\bar{x})}{a_n \frac{\p {}_1 F_1}{\p a}(a;b;\bar{y})|_{a=a_n}}
\eeq
The asymptotic of the spectrum for large $n$ is obtained from the expansion \cite{linetsky2004computing}
\beq
{}_1 F_1(a;b;z) =\pi^{-\frac{1}{2}}\Gamma(b)\exp{\frac{z}{2}} (z(b/2-a))^{\frac{1}{4}-\frac{b}{2}}\cos(2\sqrt{z(b/2-a)}-\pi b/2 +\pi/4)(1+O(|a|^{-1/2})),
\eeq
when $a$ tends to $-\infty$, $b>0$ and $z>0$. Thus,
\beq
\lambda_n \sim \frac{\pi^2 \lambda}{2\log{(\frac{1}{\alpha})}}\left(n-\frac{1}{4}\right)^2 - \lambda.
\eeq
\beq
c_n \sim \frac{(-1)^{n+1}2\pi(n-1/4)}{\pi^2(n-1/4)^2-2\log(\frac{1}{\alpha})} \sqrt{\alpha}\exp\left(\frac{\lambda}{2}\right) \left(\frac{\lambda}{\log(\frac{1}{\alpha})}\right)^{-\frac{1}{4}} \cos\left(\pi \left(n-\frac{1}{4}\right) \sqrt{\frac{\lambda}{\log(\frac{1}{\alpha})}}-\frac{\pi}{4}\right).
\eeq
The survival probability is computed from the eigenfunction expansion when the process starts at $T_{down}$. It is given by
\beq\label{HittingtimeUp}
\pP(\tau_{Up}<t) = 1 -\sum_{n=1}^{\infty} \exp(-\lambda_n t) \frac{{}_1 F_1(a_n;1;\frac{\lambda T^2_{Down}}{2\sigma})}{\frac{\p {}_1 F_1}{\p \lambda}(a_n;1;\frac{\lambda T^2_{Up}}{2\sigma})}.
\eeq
We are interested in truncating the distribution series. As the eigenvalues are increasing, we only need to compute the first two ones. For $\lambda=1$ and $\alpha=0.2$, we have $\lambda_1 \sim 0.7$ and $\lambda_2 \sim 8.4$. This computation confirms that there is a large gap difference of a factor 12 between the first two eigenvalues. Thus, we can only retain the first exponential term for the distribution of increasing time
\beq
f_{inc}(t) \sim \lambda_1(T_{Up},\lambda,\sigma) \exp(-\lambda_1(T_{Up},\lambda,\sigma) t).
\eeq
\subsection{Time distribution of spindle decreasing phase}
To estimate the time distribution of spindle decreasing phase $f_{dec}$, we start the process from $T_{Up}$ to $T_{Down}$.
The solution is the Tricomi function \cite{linetsky2004computing} and the eigenvalues satisfy the equation
\beq
U(a;b;\bar{x})=0,
\eeq
and we have
\beq
c_n = -\frac{U(a_n;b;\bar{y})}{a_n \frac{\p U}{\p a}(a;b;\bar{x})|_{a=a_n}}.
\eeq
The large-n asymptotic is
\beq
\lambda_n \sim 2\lambda \left(k_n -\frac{1}{2}\right),
\eeq
\beq
c_n \sim \frac{(-1)^{n+1}\sqrt{k_n}}{(k_n - 1/2)(\pi \sqrt{k_n}-\sqrt{\lambda})} \frac{\exp(-\frac{\lambda}{2})}{\sqrt{\alpha}} \left(\frac{\log(\frac{1}{\alpha})}{\lambda}\right)^{-\frac{1}{4}} \cos\left(2 \sqrt{k_n \log{\left(\frac{1}{\alpha}\right)}} - \pi k_n + \frac{\pi}{4}\right)
\eeq
where
\beq
k_n = n-\frac{1}{4}+\frac{2\lambda}{\pi^2} + \frac{2}{\pi} \sqrt{\left(n-\frac{1}{4}\right)\lambda} + \frac{\lambda}{\pi^2}
\eeq
To obtain an approximation of the time distribution, we compute the first two eigenvalues which give $\lambda_1 \sim 2.2$ and $\lambda_2 \sim 4.8$. Although the gap is not that large, we can consider the Poissonian distribution approximation for the decay time, associated with the probability distribution
\beq
f_{dec}(t) = \lambda_1(T_{Up},\lambda,\sigma) \exp(-\lambda_1(T_{Up},\lambda,\sigma) t).
\eeq
\subsection{Stability at the Interplay Between Deterministic and Stochastic Forces}
We now investigate the mechanisms underlying the stability and shape of spindles generated by the two-dimensional Ornstein--Uhlenbeck (OU) process. Specifically, we analyze how the interplay between deterministic drift and stochastic forcing leads to an alternation of increasing and decreasing phases in the envelope amplitude, thereby producing robust spindle-like oscillations. Our central observation is that consecutive monotonic increases or decreases in amplitude are statistically rare. Instead, the most probable trajectory involves a waxing phase followed by waning, reinforcing the notion of self-contained, transient spindles. The envelope amplitude $A(t)$ evolves according to the stochastic differential equation:
\begin{equation} \label{process}
\mathrm{d}A = -W'(A),\mathrm{d}t + \sqrt{2\sigma} , \mathrm{d}w,
\end{equation}
where the effective potential $W(A)$ is given by:
\begin{equation}
W(A) = \frac{\lambda}{2} A^2 - \sigma \log A.
\end{equation}
This potential well exhibits a unique minimum at
\begin{equation}
A_{\text{min}} = \sqrt{\frac{\sigma}{\lambda}},
\end{equation}
which corresponds to the most probable amplitude value in the stationary regime. The concavity around this minimum ensures that deviations in amplitude are typically restored, favoring oscillatory behavior. To quantify the probability of successive amplitude increases (or decreases), we discretize Eq.~\eqref{process} using an Euler--Maruyama scheme with fixed time step $\Delta t$:
\begin{equation} \label{eq_amp}
\Delta A = -W'(A) \Delta t + \sqrt{2\sigma \Delta t} , \xi,
\end{equation}
where $\xi \sim \mathcal{N}(0,1)$ is a standard Gaussian random variable. This formulation allows us to analyze the sign of amplitude increments $\Delta A$ over consecutive time steps.
We now consider two types of events:
\begin{itemize}
\item[(i)] \textbf{Consecutive increases:} Two successive steps where $\Delta A_1 > 0$ and $\Delta A_2 > 0$;
\item[(ii)] \textbf{Alternating behavior:} One increase followed by a decrease, or vice versa.
\end{itemize}
Since the deterministic drift $-W'(A)$ pulls the amplitude back toward the minimum $A_{\text{min}}$, the probability of two consecutive increases (or decreases) becomes significantly suppressed, especially when starting near $A_{\text{min}}$. This behavior creates a natural alternation between waxing and waning, stabilizing the spindle structure. We now consider the two cases where the steps are either consecutively increasing or decreasing.
\subsubsection{Asymptotic computation for the number of event sustaining an increasing step}
We consider a sequence of amplitude $A_1,...,A_n$ sampled from eq.(\ref{eq_amp}) given a starting point $A_0$. We shall compute the asymptotic of probability for an increasing sequence event $I_{n}^{+}$: "$A_{min}<A_0<...<A_n$" given by
\beq
\pP(I_{n}^{+}) = \pP\left(\bigcap_{k=0}^{n-1} A_{k+1} - A_{k}>0 \right).
\eeq
The joint probability of $n$ consecutive increments conditioned on the increment values $(A_1,..A_n)$ is given
\beq
p\left(\bigcap_{k=0}^{n-1} A_{k+1} - A_{k}>0 \mid A_1, ..., A_n \right) &=& \prod_{k=0}^{n-1} p(A_{k+1} - A_{k}>0 \mid A_{k+1}),
\eeq
where we used the independent property of the increments. Thus, using eq.(\ref{eq_amp}), we have for $k \in \{0,...,n-1\}$,
\beq \label{Relatederfc}
p(A_{k+1} - A_{k}>0 \mid A_{k+1}) = p\left(\xi > \frac{W'(A_{k+1})\sqrt{\Delta t}}{\sqrt{2\sigma}} \right)= \erfc\left(\frac{W'(A_{k+1}) \sqrt{\Delta t}}{\sqrt{2\sigma}}\right),
\eeq
where $\erfc$ is the complementary error function \cite{abramowitz1948handbook}.
In the small amplitude limit, we use the following expansion
\beq
\erfc(x) \approx \frac{\exp(-x^2)}{x \sqrt{\pi}} \hbox{ for } x \gg 1,
\eeq
thus, we have for $x=\frac{W'(A_{k+1}) \sqrt{\Delta t}}{\sqrt{2\sigma }} \gg 1$,
\beq
p(A_{k+1} - A_k>0 \mid A_{k+1}) \approx \frac{\sqrt{2\sigma}}{W'(A_{k+1}) \sqrt{\pi \Delta t }}
\exp\left(-\left(\frac{W'(A_{k+1}) \sqrt{\Delta t}}{\sqrt{2\sigma }}\right)^2\right).
\eeq
Thus the conditional probability is
\beq
p\left(\bigcap_{k=0}^{n-1} A_{k+1} - A_k>0 \mid A_1, ..., A_n \right) &\approx& \left(\frac{2\sigma}{\pi \Delta t}\right)^{\frac{n}{2}} \prod_{k=1}^{n} \frac{\exp(-\frac{\Delta t}{2\sigma} W'(A_{k})^2)}{W'(A_{k})}.
\eeq
When sampling the distance from the steady-state distribution given by eq.(\ref{pdfSteadtstate}), we have
\begin{multline}
\pP(I_{n}^{+}) = \int_{A_0}^{+\infty}...\int_{A_{n-1}}^{+\infty} p\left(\bigcap_{k=0}^{n-1} A_{k+1} - A_{k}>0 \mid A_1, ..., A_n \right) p(A_1)...p(A_n) dA_1...dA_n \nonumber \\
\approx \int_{A_0}^{+\infty}...\int_{A_{n-1}}^{+\infty} \left(\frac{\lambda}{\sigma}\right)^{n} \left(\frac{{2\sigma}}{\pi \Delta t}\right)^{\frac{n}{2}} \exp\left(-\frac{\Delta t}{2\sigma} \sum_{k=1}^{n} \left(\lambda A_k -\frac{\sigma}{A_k}\right)^2\right) \prod_{k=1}^{n} \frac{A_k \exp(-\frac{\lambda}{2\sigma} A_k^2)}{\lambda A_k -\frac{\sigma}{A_k}}  dA_1...dA_n.
\end{multline}
For $A_0$ large enough, we also have that $\lambda A_k \gg \frac{\sigma}{A_k}$ (i.e. the derivative of the potential is positive) and thus for all k$\geq$1, we have the following approximations
\beqq
\exp\left(-\frac{\Delta t}{2\sigma} \sum_{k=1}^{n} \left(\lambda A_k -\frac{\sigma}{A_k}\right)^2\right) &\approx& \exp\left(-\frac{\lambda^2 \Delta t}{2\sigma} \sum_{k=1}^{n} A_k^2\right) \\
\frac{A_k}{\lambda A_k -\frac{\sigma}{A_k}} &\approx& \frac{1}{\lambda}.
\eeqq
Thus,
\beq
\pP(I_{n}^{+}) &=& \left(\frac{2}{\pi \sigma \Delta t}\right)^{\frac{n}{2}} \int_{A_0}^{+\infty}...\int_{A_{n-1}}^{+\infty} \exp\left(-\frac{\lambda^2\Delta t}{2\sigma} \sum_{k=1}^{n} A_k^2\right) \prod_{k=1}^{n} \exp\left(-\frac{\lambda}{2\sigma} A_k^2\right) dA_1...dA_n \nonumber \\
&=& \left(\frac{{2}}{\pi \sigma \Delta t }\right)^{\frac{n}{2}} \int_{A_0}^{+\infty}...\int_{A_{n-1}}^{+\infty} \exp\left(-B \sum_{k=1}^{n} A_k^2\right) dA_1...dA_n,
\eeq
where $B = \frac{\lambda^2 \Delta t + \lambda}{2\sigma}$. The last integral can be computed by induction as
\beq
\int_{A_{n-1}}^{+\infty} \exp(-B A_n^2) dA_n = \frac{\sqrt{\pi} \erfc(\sqrt{B} A_{n-1})}{2 \sqrt{B}} \\
\int_{A_{n-2}}^{\infty} \frac{\sqrt{\pi} \erfc(\sqrt{B} A_{n-1})}{2\sqrt{B}}\exp(-B A_{n-1}^2) dA_{n-1} = \frac{\pi \erfc(\sqrt{B} A_{n-2})^2}{8 B}.
\eeq
Thus,
\beq
\int_{A_0}^{\infty}...\int_{A_{n-1}}^{\infty} \exp\left(-B \sum_{k=1}^n A_k^2\right) dA_1..dA_n = \frac{\pi^{\frac{n}{2}} \erfc(\sqrt{B} A_0)^n}{2^n n! B^{\frac{n}{2}}}.
\eeq
Finally, we conclude that the asymptotic probability of a n-consecutive increasing increment  is given
\beq
\pP(I_{n}^{+}) &\approx& \left(\frac{{2}}{\pi \sigma \Delta t}\right)^{\frac{n}{2}} \frac{\pi^{n/2} \erfc(\sqrt{B} A_0)^n}{2^n n! B^{\frac{n}{2}}} \approx \left(\frac{\erfc(\sqrt{B} A_0)}{\sqrt{2\sigma B \Delta t}}\right)^n \frac{1}{n!},
\eeq
which tends to zero as $n$ tends to infinity for any fixed $\Delta t$ and initial position $A_0>A_{min}$ (minimum of the potential well). We conclude that a consecutive increasing sequence of n terms  is  exponentially small with n, which guarrantees that the stochastic event to reach $n$ consecutive increments as $n$ increases is very unlikely.
\subsubsection{Asymptotic formula for  number of eventa sustaining a decreasing step}
In the case of a decreasing sequence, with $0<A_n<..<A_0$, a similar approach can be used to show that the asymptotic probability of such sequence becomes small as $n$ increases. Following similar steps as in the previous paragraph, the conditional probability of a decrement is
\beq
p(A_{k+1} - A_{k}<0 \mid A_{k+1}) &=& p\left(\xi_k < \frac{W'(A_{k+1})\sqrt{\Delta t}}{\sqrt{2\sigma}}\right) = \frac{1}{2}\left(1+\erf\left(\frac{W'(A_{k+1})\sqrt{\Delta t}}{\sqrt{2\sigma}}\right)\right),
\eeq
where $W'(A_{k+1}) < 0$. Thus, for $\frac{W'(A_{k+1})\sqrt{\Delta t}}{\sqrt{2\sigma}}\ll-1 $, we get the asymptotic expression
\beq
p(A_{k+1} - A_{k}<0 \mid A_{k+1}) \approx - \frac{\sqrt{\sigma}}{W'(A_{k+1}) \sqrt{2 \pi \Delta t}}
\exp\left(-\frac{\Delta t}{2\sigma} W'(A_{k+1})^2\right)
\eeq
Thus the conditional probability is now given by
\beq
p\left(\bigcap_{k=0}^{n-1} A_{k+1} - A_k<0 \mid A_1, ..., A_n \right) &=& \prod_{k=0}^{n-1} p(A_{k+1} - A_{k}<0 \mid A_{k+1}) \nonumber \\
&\approx& \left(\frac{\sigma}{2\pi\Delta t}\right)^{\frac{n}{2}} \prod_{k=1}^{n} \frac{\exp\left(-\frac{\Delta t}{2\sigma} W'(A_{k})^2 \right)}{-W'(A_{k})}.
\eeq
The asymptotic probability of $I_{n}^{-}$: "n consecutive decrements" when the distribution of distance is sampled from steady-state distributions is then given by
\begin{multline}
\pP(I_{n}^{-}) = \int_0^{A_0}...\int_0^{A_{n-1}} p\left(\bigcap_{k=0}^{n-1} A_{k+1} - A_k<0 \mid A_1, ..., A_n \right) p(A_1)..p(A_n)dA_1..dA_n \nonumber \\
\approx \left(\frac{\lambda}{\sigma}\right)^n \left(\frac{\sigma}{2\pi\Delta t}\right)^{\frac{n}{2}} \int_0^{A_0}...\int_0^{A_{n-1}} \exp\left(-\frac{\Delta t}{2\sigma} \sum_{k=1}^n \left(\lambda A_k -\frac{\sigma}{A_k}\right)^2\right) \prod_{k=1}^{n} \frac{A_k \exp\left(-\frac{\lambda}{2\sigma} A_k^2\right)}{- \lambda A_k + \frac{\sigma}{A_k}} dA_1...dA_n.
\end{multline}
For $A_0$ small enough so that $A_k \ll 1$ for all k$\geq$1, we used the following approximations
\beq
\exp\left(-\frac{\Delta t}{2\sigma} \sum_{k=1}^{n} \left(\lambda A_k -\frac{\sigma}{A_k}\right)^2\right) &\approx& \exp\left(-\frac{\sigma \Delta t}{2} \sum_{k=1}^{n} \frac{1}{A_k^2}\right) \\
\frac{A_k}{-\lambda A_k + \frac{\sigma}{A_k}} &\approx& \frac{A^2_k}{\sigma},
\eeq
We set $C=\frac{\sigma \Delta t}{2}$ and integrate by parts the last integral giving
\beq
\int_0^{A_{n-1}} \exp\left(-\frac{C}{A_n^2}\right) A_n^2 dA_n &=& C^{\frac{3}{2}}\int_{\frac{\sqrt{C}}{A_{n-1}}}^{\infty} \frac{\exp(-u^2)}{u^4} du \nonumber \\
&=& \frac{1}{2C}\exp\left(-\frac{C}{A_{n-1}^2}\right) A_{n-1}^5, \\
\frac{1}{2C} \int_0^{A_{n-2}} \exp\left(-\frac{2C}{A_{n-1}^2}\right) A_{n-1}^7 dA_{n-1} &=& (2C)^3\int_{\frac{\sqrt{2C}}{A_{n-2}}}^{\infty} \frac{\exp(-u^2)}{u^9} du \nonumber \\
&=& \frac{1}{2 (2C)^2} \exp\left(-\frac{2C}{A_{n-2}^2}\right) A_{n-2}^{10}.
\eeq
Thus, we obtain by induction the general formula
\beq
\int_0^{A_0}...\int_0^{A_{n-1}} \prod_{k=1}^{n} \exp\left(-\frac{\sigma \Delta t}{2A_k^2}\right) A_k^2 dA_k = \frac{1}{2^n n!C^n} \exp\left(-\frac{nC}{A_0^2}\right) A_{0}^{5n}.
\eeq
Finally,
\beq
\pP(I_{n}^{-}) &\approx& \left(\frac{\sigma}{2\pi\Delta t}\right)^{\frac{n}{2}} \left(\frac{\lambda}{\sigma^3 \Delta t}\right)^n \frac{1}{n!} \exp\left(-\frac{n\sigma \Delta t}{2A_0^2}\right) A_0^{5n}.
\eeq
which tends to zero as $n$ increases. To conclude, since the asymptotic probability to have a strictly increasing or decreasing sequence becomes small as $n$ increases, most events consist of a few consecutive increasing and decreasing epochs, thus preventing the spindle envelope to grow infinitely. It is interesting to note the asymmetry between the increasing and decreasing cases.
\subsection{Mean number of consecutive increments}
To determine the mean time of an increasing sequence starting at an amplitude $A_0$, we  use a mean first passage approach to define the last increasing step as
\beq
\tilde n = \inf\{ n \hbox { such that } A_{q+1}-A_{q}> 0, q\in\{0,..,n-1\} \hbox{ and } A_{n+1}-A_{n}< 0\}.
\eeq
We are interested in computing the expectation
\beq
\eE(\tilde n) =\sum_{n=1}^{\infty} n \pP(\tilde n=n),
\eeq
where $\pP(\tilde n=n)$ is the probability that there are exactly $n$ consecutive increments. We can express the probability to have exactly n jumps as the average over all configuration where $A_{q+1}>A_q$, for all possibility when starting at $A_{0}>A_{min}$ given the steady-state
\begin{multline}
\pP(\tilde n=n) = \int_{A_0}^{\infty}...\int_{A_{n-1}}^{\infty}\int_{0}^{A_{n}} p(A_{n+1}-A_{n}<0 \mid A_{n}) p(A_{n+1}) \prod_{k=1}^{n} p(A_{k+1}-A_{k}>0 \mid A_{k}) p(A_{k}) dA_1...dA_{n+1}.
\end{multline}
\beq
\int_{0}^{A_{n}} p(A_{n+1}) p(A_{n+1}-A_{n}>0 \mid A_{n})dA_{n+1}=I_1+I_2(A_n)
\eeq
where
\beq
I_1&=&\int_{0}^{A_{min}} p(A_{n+1}-A_{n}>0 \mid A_{n})p(A_{n+1}) dA_{n+1}\\
I_2(A_n)&=&\int_{A_{min}}^{A_n}  p(A_{n+1}-A_{n}>0 \mid A_{n})p(A_{n+1}) dA_{n+1}.
\eeq
In the first interval, the potential is negative and the other one positive.
\beq
I_1(A_n)=\frac{\lambda}{\sigma} \int_{0}^{A_{min}} p\left(\xi_k < \frac{W'(A_{k+1})\sqrt{\Delta t}}{\sqrt{2\sigma}}\right)
A_{n+1} \exp(-\frac{\lambda}{2\sigma} A_{n+1}^2)dA_{n+1}.
\eeq
In the interval $[0,A_{min}]$, $W'<0$. We will make the approximation of $W'(A)\sim -\frac{\sigma}{A}$ so that
\beq
I_1(A_n) &=&\left(\frac{\sigma}{2\pi\Delta t}\right)^{\frac{1}{2}}
\frac{\lambda}{\sigma} \int_{0}^{A_{min}} \exp\left(-\frac{\sigma \Delta t}{2A_n^2}\right) A_{n+1}^2 \exp(-\frac{\lambda}{2\sigma} A_{n+1}^2)dA_{n+1} \\&=&\left(\frac{\sigma}{2\pi\Delta t}\right)^{\frac{1}{2}}
\frac{\lambda}{\sigma} \int_{0}^{A_{min}} \exp\left(-\frac{\sigma \Delta t}{2A_n^2}\right) A_{n+1}^2(1+O(1))dA_{n+1} \\
&=&
\left(\frac{\sigma}{2\pi\Delta t}\right)^{\frac{1}{2}}
\frac{\lambda}{\sigma^2 \Delta t}\exp\left(-\frac{\sigma}{2\pi\Delta t A_{min}^2}\right) A_{min}^5
\eeq
Similarly,
\beqq
I_2(A_n)&=&\int_{A_{min}}^{A_n}  p(A_{n+1}-A_{n}>0 \mid A_{n})p(A_{n+1}) dA_{n+1}\\&=&\frac{\lambda}{\sigma}
\int_{A_{min}}^{A_n}  p\left(\xi_k < \frac{W'(A_{k+1})\sqrt{\Delta t}}{\sqrt{2\sigma}}\right)
A_{n+1} \exp(-\frac{\lambda}{2\sigma} A_{n+1}^2)dA_{n+1}\\&=& \frac{\lambda}{\sigma}
\int_{A_{min}}^{A_n} \frac{1}{2}\left(1+\erf\left(\frac{W'(A_{k+1})\sqrt{\Delta t}}{\sqrt{2\sigma}}\right)\right) A_{n+1} \exp(-\frac{\lambda}{2\sigma} A_{n+1}^2)dA_{n+1}=\\
&=&\exp (-\frac{\lambda}{2\sigma} A_{min}^2) -\exp (-\frac{\lambda}{2\sigma} A_{n}^2)\approx \exp (-\frac{\lambda}{2\sigma} A_{min}^2)
\eeqq
where we approximated the erf function to 1.  Thus we obtain for the integral:

\begin{multline}
\pP(\tilde n=n) = \int_{A_0}^{\infty}...\int_{A_{n-1}}^{\infty}  (C_{\min}
\prod_{k=1}^{n} p(A_{k+1}-A_{k}>0 \mid A_{k}) p(A_{k}) dA_1...dA_{n+1},
\end{multline}
where
$C_{\min}=
\left(\frac{\sigma}{2\pi\Delta t}\right)^{\frac{1}{2}}
\frac{\lambda}{\sigma^2 \Delta t}\exp\left(-\frac{\sigma}{2\pi\Delta t A_{min}^2}\right) A_{min}^5+ (\exp (-\frac{\lambda}{2\sigma} A_{min}^2).
$
Finally,
\beq
\pP(\tilde n=n) = C_{\min} \left(\frac{\erfc(\sqrt{B} A_0)}{\sqrt{2\sigma B \Delta t}}\right)^n \frac{1}{n!}.
\eeq
We conclude that the mean number of increments is given by
\beq
\eE(\tilde n|A_0) \approx  C_{\min} (\exp (\frac{\erfc(\sqrt{B} A_0)}{\sqrt{2\sigma B \Delta t}})-1)\frac{\erfc(\sqrt{B} A_0)}{\sqrt{2\sigma B \Delta t}},
\eeq
where we recall $B = \frac{\lambda^2 \Delta t + \lambda}{2\sigma}$.
Finally, it would be interesting to compute the number of decreasing steps as:
\beq
\tilde m = \inf\{ m \hbox { such that } A_{m+1}-A_{m}> 0 \hbox{ and } A_{q+1}-A_{q}< 0, q\in\{0,..,n-1\}\}.
\eeq
the mean number of steps is given by the expectation $\eE(\tilde m) =\sum_{m=1}^{\infty} m \pP(\tilde m=m)$.
\subsection{A general formula for the distribution of spindle amplitude}
We propose here to derive a general expression for the distribution of maximum spindle amplitude. This is in principle an uneasy task to compute as computing such a distribution requires segmenting each spindle and identifying the maximum. Instead we shall compute the distribution of maximum of the radius in a spindle time interval. Due to the independent property of spindles, we will be interested in computing the probability that the maximum hits level u when starting at height $x$
\beq
\pP\left(\max_{s\in[0,t]} A(s)=u \mid x>0\right).
\eeq
When conditioning this  process on the duration $t_s$ of the spindle, we get
\beq
\pP\left(\max_{s\in[0,t_s]} A(s)=u \mid t_s=t,x\right).
\eeq
so that by Bayes' law,
\beq\label{maxdistrib}
\pP\left(A_{max}=u\right)=\int_0^{\infty} \pP\left(\max_{s\in[0,t_s]} A(s)=u \mid t_s=t,x\right) \pP\left( t_s=t \right)dt,
\eeq
where $\pP\left( t_s=t \right)$ is the probability that the duration of a spindle is equal to t, that we show to be a degenerated difference of Poisson process given by
\beq
\pP\left( t_s=t \right)=\lambda^2 t \exp(-\lambda t),
\eeq
where we consider here that the mean increasing and decreasing time are identical, thus $\lambda=\lambda_{inc}=\lambda_{dec}$. Moreover, the probability $\pP\left(\max_{s\in[0,t_s]} A(s)<u \mid t_s=t\right)$ is the survival of the process inside the interval $[0,u]$ and can  be computed from the backward Fokker-Planck equation
\beq\label{FPE2}
\frac{\p}{\p t}p(x,t) &=& \frac{\epsilon x}{2} \frac{\p^2}{\p x^2} p(x,t) + \kappa(\theta-x)\frac{\p }{\p x}p(x,t) \nonumber \\
p(x,0) &=& \delta(x-y) \\
p(u,t) &=& 0. \nonumber
\eeq
We recall that this process has not reached the level $u$ is equivalent to the hitting time being less than $t$ when the process starts at $x$, thus
\beq
\pP(\tau^u<t \mid x) = \pP\left(\max_{s\in[0,t_s]} A(s)\leq u \mid t_s=t,x \right).
\eeq
We get directly from eq.(\ref{HittingtimeUp}) the pdf of the time, condition on the initial value $x$
\beq
\pP(\tau^u=t \mid x) = \kappa\sum_{n=1}^{\infty} \exp(-\lambda_n(u) t) \frac{ {}_1 F_1(a_n;1;\frac{2\kappa u}{\eps} x)}{\frac{\p {}_1 F_1}{\p a}(a;1;\frac{2 \kappa u}{\eps})|_{a=a_n}}.
\eeq
where the spectrum $\lambda_n(u) $ is given by the transcendental equation with $a_n = -\frac{\lambda_n}{\kappa}$
\beq
{}_1 F_1(a_n;1;\frac{2\kappa u}{\eps})=0,
\eeq
leading to $\lambda_n(u)$, n=1...
Thus, we obtain from relation (\ref{maxdistrib}),
\beq \label{maxdistrib2}
\pP(A_{max} =u \mid x) &=& \int_0^{\infty} \pP\left(\max_{s\in[0,t_s]} A(s)=u \mid t_s=t\right) \pP\left( t_s=t \right) dt \nonumber \\
&=& \sum_{i} \frac{\lambda^2}{(\lambda_n(u)+\lambda)^2} \frac{ {}_1 F_1(a_n;1;\frac{2\kappa u}{\eps} x)}{\frac{\p {}_1 F_1}{\p a}(a;1;\frac{2 \kappa u}{\eps})|_{a=a_n}}.
\eeq
It would be interesting to obtain a simplified expression and a final expression for the probability of the maximum with respect to the parameter of the Ornstein-Ulhembeck process.
\section{Statistical properties of spindles generated by the OU process}
\begin{figure*}[http!]
\centering
\includegraphics[width=0.85\linewidth]{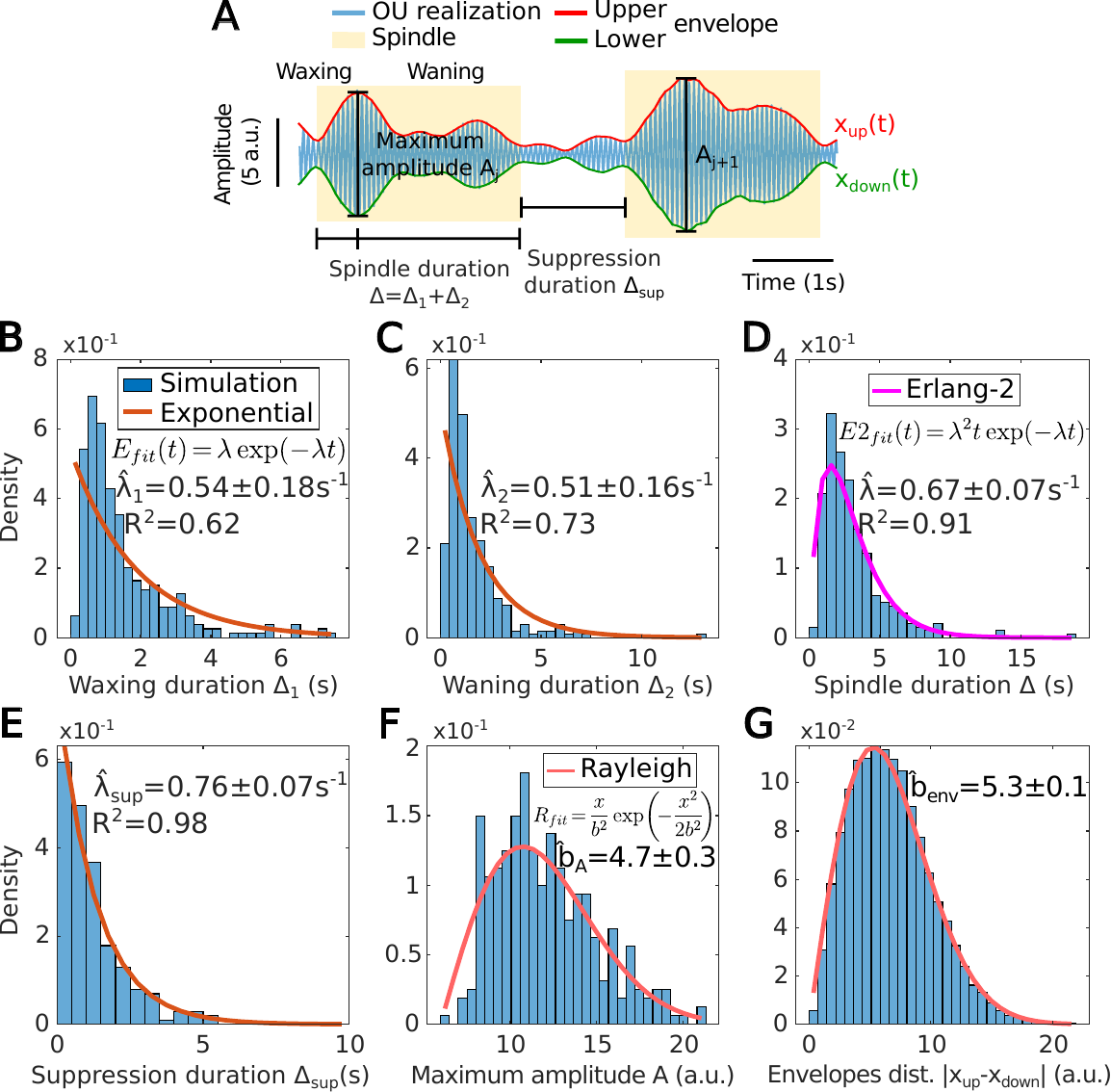}
\caption{{\bf Statistics of spindles generated by an OU process.}
{\small
{\bf (A)} Segmentation of spindles generated by stochastic simulation of an OU process with parameters $(\lambda,\omega,\sigma) = (1,60,8)$. The spindle starts when the envelope $x_{up}$ exceeds for the first time the threshold $T$ and ends when it return to $T$. The total spindle duration $\Delta$ is the sum of duration associated to the increasing phase $\Delta_1$ and the decreasing phase $\Delta_2$.
{\bf (B)} Distribution of spindle increasing phase duration $\Delta_1$, fitted with an exponential distribution of rate $\hat{\lambda}_1=0.54\pm0.18 s^{-1}$ and $R^2=0.62$.
{\bf (C)} Distribution of spindles decreasing phase duration $\Delta_2$, with an exponential distribution of rate $\hat{\lambda}_2=0.51\pm0.16 s^{-1}$ and $R^2=0.73$.
{\bf (D)} Distribution of total spindle duration $\Delta$, with an Erlang-2 distribution of rate $\hat{\lambda}_1=0.54\pm0.18 s^{-1}$ and $R^2=0.62$.
{\bf (E)} Distribution of suppression duration $\Delta_{sup}$ (region between two successive spindles).
{\bf (F)} Distribution of the spindle maximum amplitude $A$, fitted with a Rayleigh distribution of scale $b_A=4.7\pm0.3$.
{\bf (G)} Distribution of the envelope amplitude $|x_{up}-x_{down}|$, fitted with a Rayleigh distribution of scale $b_{env}=5.3\pm0.1$.
}}
\label{fig:5}
\end{figure*}

In this section, we analyze the statistical properties of spindles generated by the two-dimensional OU-process that we first segmented using the method presented above (section \ref{s:segmentation}). our goal here is to determine the distribution of their total duration, as well as the one for the increasing phase (waxing) and decreasing phase (waning). The result are presented in Fig. \ref{fig:5}.\\
We recall the two phases: the increasing phase starts at a time $t_j$ and ends when the process has reached a maximum amplitude $A_j$, lasting $\Delta_1$. This phase is followed by a decreasing phase starting from the maximum to the next minimum $t_{j+1}$, lasting $\Delta_2$ (Fig. \ref{fig:5}A). The total spindle duration is thus the sum $\Delta = \Delta_1 + \Delta_2$. \\
The statistics revealed from the stochastic simulations are presented in Fig. \ref{fig:5}B-C, where the increasing and decreasing phase duration distributions can be each well approximated by a single exponential curve (as discussed in section \ref{s:waxing}). We thus fitted the exponential function $f_{\Delta_1}(t) = \lambda_1 \exp(-\lambda_1 t)$ and $f_{\Delta_2}(t) = \lambda_2 \exp(-\lambda_2 t)$, where the optimal fits lead to $\lambda_1=0.54\pm 0.18 s^{-1}$ ($R^2=0.62$) and $\lambda_2=0.51\pm0.16 s^{-1}$ ($R^2=0.73$). Finally, we fitted the spindle total duration by an Erlang-2 distribution $f_{\Delta}(t) = \lambda^2 t \exp(-\lambda t)$, this fit is based on the result that the increasing and decreasing phase duration are quite similar, as we found above. The optimal fit (Fig. \ref{fig:5}D) yielded $\lambda=0.67\pm 0.07 s^{-1}$ ($R^2=0.91$). \\
We further investigated the distribution of suppression duration $\Delta_{sup}$, which is defined as the absence of spindle above the threshold $T_{down}$. We fitted a single exponential \cite{schuss2009theory} leading to the estimated rate constant $\lambda_{sup} = 0.76\pm 0.07 s^{-1}$ (Fig. \ref{fig:5}E). Interestingly, this mean expectation is in agreement with the estimation of the mean first passage time that we derived in subsection \ref{interspindleestimation}. \\
Finally, the distribution of spindles maximum amplitude $A$ computed over each spindle (Fig. \ref{fig:5}F) can be fitted by a Rayleigh distribution, which is the steady-state distribution that we obtained from the solution of the Fokker-Plank equation (\ref{pdfSteadtstate}). The Rayleigh distribution is the solution of the distribution of the maximum of the envelope distance $|x_{up}-x_{down}|$ that we computed from the simulation of the stochastic eq. \ref{ou_eq}, where the optimal fits lead to $b_A=4.7\pm0.3$ and $b_{env}=5.3\pm0.1$. \\
To conclude, we showed that the spindle
generated by the OU process can be segmented using the increasing and decreasing phase of the envelope. The increasing phase can be interpreted as driven by noise to a random maximum distance of the variable $\s$, while the decreasing phase could be interpreted as a relaxation to a region where the amplitude size reaches the lower threshold $T_{down}$. We showed that the duration of increasing and decreasing phases are well approximated by single exponential distribution. We proposed a series formula for the distribution of the spindle maximum amplitude, but we could not find a closed form. However, the Rayleigh distribution seemed to be a good approximation.
\section{Coupling Two OU Processes to Generate Spindles of Mixed Frequencies}
\begin{figure}[http!]
\centering
\includegraphics[width=1\linewidth]{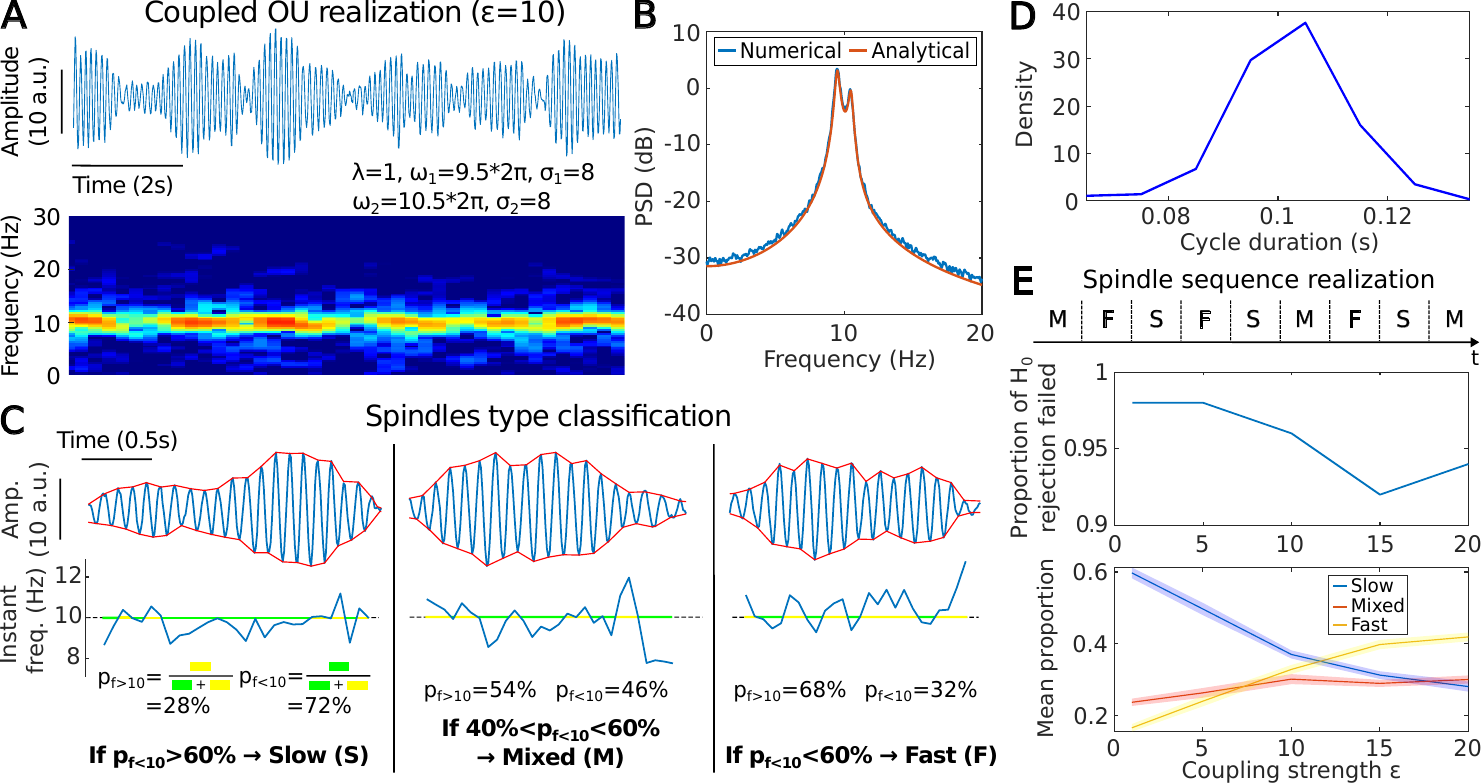}
\caption{{\bf Spindles classification and comparison in the coupled OU model.}
{\small
{\bf (A)} Example of a coupled OU realization with $\eps$=10 (top) with its corresponding spectrogram (bottom).
{\bf (B)} Numerical (blue) and theoretical (red) PSD of the coupled OU realization shown in (A).
{\bf (C)} Three examples of spindles (top) and their instantaneous frequency across time (bottom). The proportion of slow/fast oscillations is computed as the ratio between duration for which the instantaneous frequency is respectively below/above (green/yellow) 10 Hz and the spindle total duration. When the proportion of slow (resp. fast) oscillations is greater than 60\%, the spindle is classified as slow (resp. fast). If both proportions are between 40\% and 60\%, it is classified as mixed.
{\bf (D)} Mean density of spindle cycle duration.
{\bf (E)} Top: Schematic of spindle sequence. Middle: To test whether spindles appear in random or deterministic order, we varied $\eps$ and runs test on 50 simulations for each choice of $\eps$ and calculated the proportion of trials where the null hypothesis (H0: the sequence is deterministic) was not rejected. Bottom: Evolution of slow (blue), mixed (red) and fast (yellow) spindle proportion with respect to the coupling strength $\eps$. The shaded area represents the standard error.
}}
\label{fig:6}
\end{figure}

To account for the diversity of spindle types observed in neural recordings—particularly with respect to their frequency content—we consider coupling multiple two-dimensional Ornstein--Uhlenbeck (OU) processes. In this section, we analyze how coupling two such processes, each tuned to a distinct  frequency, can generate spindles exhibiting multiple dominant frequency bands.
Specifically, we couple two two-dimensional stochastic foci with resonance frequencies  $\omega_1=2\pi f_1$ and $\omega_2=2\pi f_2$. The coupling is made by a small parameter $\epsilon$ as followed:
\beq
\Dot{X_1} &=& -\lambda X_1 + \omega_1 Y_1 + \epsilon X_2 + \sqrt{2\sigma_1} \Dot{w_1} \nonumber \\
\Dot{Y_1} &=& -\omega_1 X_1 - \lambda Y_1 + \sqrt{2\sigma_1} \Dot{w_2} \\
\Dot{X_2} &=& -\lambda X_2 + \omega_2 Y_2 + \sqrt{2\sigma_2} \Dot{w_3} \nonumber \\
\Dot{Y_2} &=& -\omega_2 X_2 - \lambda Y_2 + \sqrt{2\sigma_2} \Dot{w_4}, \nonumber
\eeq
where $w_i$ are Brownian motions and $\lambda$ is the decaying parameter. \\
A representative realization of this coupled system is shown in Fig.\ref{fig:6}A), and the PSD of the first component $x_1$ shows two peaks (Fig. \ref{fig:6}B), that we fitted using an exact computation (Appendix B). This result confirms that the coupling leads to various spindle types and that two frequencies are present.

At this stage, we identified three types of spindles as shown in Fig. \ref{fig:6}C-E: slow (S) ($\text{min}(f_1,f_2)$), fast (F) ($\text{max}(f_1,f_2)$) and mixed (M) which contain approximately the same proportion of both frequencies. We classified spindles according to the following criteria. We label a spindle slow when more than 60\% of the instantaneous frequency is less than 10 Hz, fast when more than 60\% of the instantaneous frequency is greater than 10 Hz. Finally, it is considered mixed when both slow and fast frequency proportion is between 40\% and 60\% (Fig. \ref{fig:6}C). \\
Moreover, to investigate whether a spindle sequence could carry information (Fig. \ref{fig:6}E), we estimated the spindle type sequence for a fixed value of $\eps$.  To evaluate the randomness of the three-valued $(S,F,M)$ spindle sequence, we use the Wald-Wolfowitz runs test \cite{bradley1960distribution,sheskin2003handbook}. A run is defined as a consecutive sequence of identical values. The null hypothesis $H_0$ assumes that the sequence is random. The number of runs under H0 can be approximated by the expected number of runs given by
\beq
\eE(R) = \frac{2n_S n_F}{n} + \frac{2n_S n_M}{n} + \frac{2n_F n_M}{n} + 1
\eeq
where $n_S$, $n_F$, and $n_M$ are the counts of the slow, fast and mixed spindles and $n = n_A + n_B + n_C$ is the total number of spindles. The variance of the number of runs is calculated as
\beq
\Var(R) = \frac{2n_S n_F (2n_S n_F - n)}{n^2 (n - 1)} + \frac{2n_S n_M (2n_S n_M - n)}{n^2 (n - 1)} + \frac{2n_F n_M (2n_F n_M - n)}{n^2 (n - 1)},
\eeq
and the test statistic $Z$ is
\beq
Z = \frac{R - E(R)}{\sqrt{\text{Var}(R)}}
\eeq
where $R$ is the observed number of runs. The value of $Z$ is compared to critical values from the standard normal distribution. If $|Z|$ exceeds the critical value 1.96 for $\alpha = 0.05$, H0 is rejected, indicating that the sequence is not random.\\
We performed the runs test on 50 coupled OU process simulations of duration 20 minutes. We found that at least 90\% of trials were not rejected for a large range of $\eps$ values (Fig. \ref{fig:6}E upper). These findings support the hypothesis that the sequence of spindle types—slow, fast, and mixed—emerges in a statistically random order for a broad range of coupling strengths $\epsilon$ ans thus no physiological information should be expected in the EEG signal when extracting the spindle order. \\
Finally, we calculated the proportion of each spindle type $p_A$ for $A\in{S,F,M}$ as the ratio $n_A/n$. By running simulations with varying values of the coupling strength $\epsilon$, we found that the proportion of slow, mixed and fast spindles depends strongly on $\eps$ (Fig. \ref{fig:6}E lower). Indeed, the proportion of slow spindles significantly decreased when the coupling strength $\eps$ increases. \\
To conclude,  by systematically varying the coupling strength $\\epsilon$, we observed a pronounced shift in the distribution of spindle types. In the weak coupling regime ($\epsilon \approx 0$), spindles were primarily slow, reflecting the dominance of the lower frequency oscillator. As $\epsilon$ increased, the fraction of mixed-frequency spindles grew, while the proportion of slow spindles declined markedly. This transition illustrates how increasing interaction between the two oscillators allows the higher frequency component to influence the output, thereby broadening the frequency content of individual spindles. At strong coupling, fast spindles dominate, confirming that coupling strength modulates the richness of frequency expression in the model. Interesting, mixed spindles seems to saturate already for $\epsilon=10$, suggesting  by further increasing $\eps$ could have no further effect.\\
These results suggest a potential mechanism for frequency blending in biological systems, where weakly interacting oscillatory populations can give rise to diverse, stochastic, yet physiologically plausible spindle events.
To conclude, this analysis revealed that coupling stochastic focus can generate randomly a variety of spindles containing multiple frequencies, however, we expect no specific physiological significance of their order based on the present stochastic model. It would be interesting to generalize the present model by adding a time dependent coupling that could reflect the possible change of interaction between different frequency bands.
\section{Discussion and Open Problems}
In this work, we presented a minimal yet analytically tractable stochastic model for generating spindle-like events—transient bursts of oscillatory activity characterized by distinct waxing (increasing) and waning (decreasing) phases. We showed that such events  emerge from a two-dimensional Ornstein–Uhlenbeck (OU) process, where rotational drift and stochastic forcing interact to produce repeated excursions from a stable focus. We developed a  method to segment these spindles and  characterize their statistical properties, including the durations of their ascent and descent phases, the amplitude of peak oscillations, and the inter-spindle intervals.

We also extended the framework by coupling multiple OU processes with different intrinsic frequencies. This leads to the emergence of composite or mixed-frequency spindles, providing a mechanism to model more complex temporal patterns observed in empirical data.

Our model exhibits striking similarities to spindles found in electroencephalogram (EEG) recordings, particularly in the $\alpha$-band (8--12 Hz), which is prominent during general anesthesia, stage 2 non-REM (NREM) sleep, quiet wakefulness, and meditative states \cite{lopes1974model}. Traditionally, the $\alpha$-band appears in spectrograms as a continuous frequency trace, but closer inspection—such as illustrated in Fig.~\ref{fig:1}A--B—reveals it is composed of successive, non-overlapping spindle events.

The current framework can be further generalized to model richer EEG spindle phenomena. For instance, by incorporating multiple coupled oscillatory modes, one may explore the coexistence and interaction of fast ($>$13 Hz) and slow ($<$13 Hz) spindles, which are notably observed during stage 2 NREM sleep \cite{Molle2011}. These spindles are region-specific—slow spindles are more prominent in frontal cortical regions, while fast spindles tend to occur more posteriorly—and are believed to result from interactions between the thalamic reticular nucleus, the thalamus, and the neocortex. Functionally, spindles have been implicated in sensory disconnection during sleep, consolidation of learning and memory, and motor planning.

Despite the progress offered by our model, several computational and segmentation  questions remain open:
\begin{enumerate}
    \item \textbf{Spindle statistics as biomarkers:} Can deviations in the statistical distributions of spindle properties (e.g., durations, amplitudes, inter-spindle intervals) indicate physiological transitions, such as the onset of sleep stages, emergence from anesthesia, or pathological states?

    \item \textbf{Spectral bandwidth from coupling:} How does coupling between multiple OU processes affect the effective width and shape of the frequency spectrum? Can one derive closed-form expressions for the power spectral density and its bandwidth in the case of arbitrary coupling?

    \item \textbf{Distribution of spindle amplitudes:} While we have characterized mean amplitudes, is there a closed-form expression (or asymptotic estimate) for the full probability distribution of the maximum spindle amplitude, and how does it depend on system parameters such as noise intensity and damping?

    \item \textbf{Detection robustness:} How sensitive is the EMD-based segmentation algorithm to changes in noise level, sampling rate, and the presence of other oscillatory components? Can the algorithm be adapted for real-time detection or used in closed-loop control systems?

    \item \textbf{Extension to non-Gaussian noise:} Can this framework be generalized to include non-Gaussian perturbations (e.g., Lévy noise or state-dependent diffusion) to model heavy-tailed statistics observed in certain neural systems?
\end{enumerate}
Addressing these problems would deepen our understanding of the stochastic origins of transient neural oscillations and may open new avenues for modeling, detection, and control in neuroscience and clinical engineering.
\section*{Acknowledgments}
D.H. research is supported by ANR AnalysisSpectralEEG, AstroXcite and Memolife, European Research Council (ERC) under the European Union’s Horizon 2020 research and innovation programme (grant agreement No 882673).
\section*{Competing interests}
The Authors declare no competing financial interests.

\newpage
\section*{Appendix A: Autocorrelation Function of the OU Process} \label{section:autocorr}
In this appendix, we compute the autocorrelation function $C_{xx}(t, s)$ for the first component $x(t)$ of the two-dimensional Ornstein–Uhlenbeck (OU) process. The stochastic differential equation is given by
\begin{equation} \label{eqfdt2}
\dot{\boldsymbol{s}} = A \boldsymbol{s} + \Sigma \, \dot{\boldsymbol{\omega}},
\end{equation}
where $\boldsymbol{s} = (x, y)^T \in \mathbb{R}^2$ is the state vector, $A$ is the drift matrix (defined in Eq.~\eqref{ou_eq}), and $\Sigma = \sqrt{2\sigma} \, I_2$ is the noise intensity matrix. The noise term $\dot{\boldsymbol{\omega}} = (\dot{\omega}_1, \dot{\omega}_2)^T$ consists of two independent white noise processes. Assuming the process has zero mean $\mathbb{E}[x(t)] = 0$, the autocorrelation function is defined as \cite{gardiner1985handbook,schuss2009theory}
\begin{equation}
C_{xx}(t, s) = \mathbb{E}[x(t) x(s)].
\end{equation}
The solution to Eq.~\eqref{eqfdt2} can be written as
\begin{equation}
\boldsymbol{s}(t) = \exp(tA) \boldsymbol{s}_0 + \int_0^t \exp((t - s)A) \Sigma \, d\boldsymbol{\omega}_s,
\end{equation}
where the second term is a stochastic convolution. Taking expectations, we obtain $\mathbb{E}[\boldsymbol{s}(t)] = \exp(tA)\boldsymbol{s}_0$, which vanishes in the stationary limit. The covariance matrix of the process is given by
\begin{equation}
C(t, s) = \mathbb{E}\left[\int_0^t \int_0^s \exp((t - s_2)A) \Sigma \Sigma^T \exp((s - s_1)A^T) \, d\boldsymbol{\omega}(s_1) \, d\boldsymbol{\omega}(s_2)\right].
\end{equation}
Using the $\delta$-correlation property of white noise, this simplifies to
\begin{align}
C(t, s) &= \int_{-\infty}^{\min(t, s)} \exp(A(t - t')) \Sigma \Sigma^T \exp(A^T(s - t')) \, dt' \\
        &= 2\sigma \int_{-\infty}^{\min(t, s)} \exp(A(t - t')) \exp(A^T(s - t')) \, dt'. \nonumber
\end{align}
Next, we decompose the drift matrix $A$ as
\begin{equation}
A = -\lambda I_2 + \omega J_2, \quad \text{where} \quad J_2 =
\begin{pmatrix}
0 & 1 \\
-1 & 0 \\
\end{pmatrix},
\end{equation}
so that the exponential of $A$ is given by
\begin{align}
\exp(tA) &= e^{-\lambda t} \left[\cos(\omega t) I_2 + \sin(\omega t) J_2\right], \\
\exp(tA^T) &= e^{-\lambda t} \left[\cos(\omega t) I_2 + \sin(\omega t) J_2^T\right].
\end{align}
Using $\tau = t - s$, we distinguish two cases.
\paragraph{Case 1: $\tau > 0$.}
Then, the correlation becomes
\begin{align}
C(t, s) &= 2\sigma \int_{-\infty}^{s} e^{-2\lambda(s - t')} e^{-\lambda \tau}
\begin{pmatrix}
\cos(\omega \tau) & -\sin(\omega \tau) \\
\sin(\omega \tau) & \cos(\omega \tau) \\
\end{pmatrix} dt' \\
&= \frac{\sigma}{\lambda} e^{-\lambda \tau}
\begin{pmatrix}
\cos(\omega \tau) & -\sin(\omega \tau) \\
\sin(\omega \tau) & \cos(\omega \tau) \\
\end{pmatrix}.
\end{align}

\paragraph{Case 2: $\tau < 0$.}
Similarly, we obtain
\begin{align}
C(t, s) &= 2\sigma \int_{-\infty}^{t} e^{-2\lambda(t - t')} e^{\lambda \tau}
\begin{pmatrix}
\cos(\omega \tau) & -\sin(\omega \tau) \\
\sin(\omega \tau) & \cos(\omega \tau) \\
\end{pmatrix} dt' \\
&= \frac{\sigma}{\lambda} e^{\lambda \tau}
\begin{pmatrix}
\cos(\omega \tau) & -\sin(\omega \tau) \\
\sin(\omega \tau) & \cos(\omega \tau) \\
\end{pmatrix}.
\end{align}
Combining both cases, the autocorrelation matrix for any $\tau \in \mathbb{R}$ becomes
\begin{equation}
C(\tau) = \frac{\sigma}{\lambda} e^{-\lambda |\tau|}
\begin{pmatrix}
\cos(\omega \tau) & -\sin(\omega \tau) \\
\sin(\omega \tau) & \cos(\omega \tau) \\
\end{pmatrix}.
\end{equation}

\paragraph{Component-wise autocorrelation.}
Extracting the $(1,1)$ component, we obtain the autocorrelation function for the first coordinate $x(t)$:
\begin{equation} \label{correlationexpression}
C_{xx}(\tau) = \frac{\sigma}{\lambda} e^{-\lambda |\tau|} \cos(\omega \tau).
\end{equation}

\paragraph{Power spectral density.}
Taking the Fourier transform of the autocorrelation function \eqref{correlationexpression}, we obtain the power spectral density (PSD) of $x(t)$:
\begin{equation}
S(f) = \frac{\sigma}{2\pi} \left( \frac{1}{\lambda^2 + (\omega - 2\pi f)^2} + \frac{1}{\lambda^2 + (\omega + 2\pi f)^2} \right).
\end{equation}
This expression reveals that the power spectrum is sharply peaked at the resonant frequency $f = \omega / 2\pi$, and its width is governed by the damping parameter $\lambda$. This relationship can be exploited to estimate $\lambda$ from data, once $\omega$ and $\sigma$ are known or independently estimated.
\newpage
\section*{Appendix B: Correlation Function and Power Spectrum for Coupled OU Processes}
In this appendix, we compute the autocorrelation function and the power spectral density (PSD) of the first variable $X_1(t)$ for a system of two weakly coupled two-dimensional Ornstein--Uhlenbeck (OU) processes. Each OU process is characterized by its own intrinsic angular frequency $\omega_1$ or $\omega_2$, and both are subject to a common damping rate $\lambda$. The coupling strength between the two systems is denoted by $\epsilon$. The noise intensities for the two processes are $\sigma_1$ and $\sigma_2$ respectively.

\paragraph{System of Equations.}
The coupled system is described by the following stochastic differential equations:
\begin{align}
\dot{X}_1 &= -\lambda X_1 + \omega_1 Y_1 + \epsilon X_2 + \sqrt{2\sigma_1} \, \dot{w}_1, \\
\dot{Y}_1 &= -\omega_1 X_1 - \lambda Y_1 + \sqrt{2\sigma_1} \, \dot{w}_2, \\
\dot{X}_2 &= -\lambda X_2 + \omega_2 Y_2 + \sqrt{2\sigma_2} \, \dot{w}_3, \\
\dot{Y}_2 &= -\omega_2 X_2 - \lambda Y_2 + \sqrt{2\sigma_2} \, \dot{w}_4,
\end{align}
where $(X_1, Y_1)$ and $(X_2, Y_2)$ are the state variables of the first and second OU processes, respectively, and $\dot{w}_i$ are independent standard white noises. The damping rate $\lambda > 0$ is shared across both subsystems.
\paragraph{Matrix Formulation.}
This system can be written compactly as:
\begin{equation}
\dot{\mathbf{Z}} = A \mathbf{Z} + \sqrt{2}\, \boldsymbol{\Sigma} \, \dot{\mathbf{W}},
\end{equation}
where $\mathbf{Z} = (X_1, Y_1, X_2, Y_2)^T$, $\dot{\mathbf{W}} = (\dot{w}_1, \dot{w}_2, \dot{w}_3, \dot{w}_4)^T$, and the drift matrix $A$ is given by:
\begin{equation}
A = -\lambda I_4 + J[\omega_1, \omega_2] + \epsilon E_{13},
\end{equation}
with the rotational and coupling matrices defined by:
\[
J[\omega_1, \omega_2] =
\begin{pmatrix}
0 & \omega_1 & 0 & 0 \\
-\omega_1 & 0 & 0 & 0 \\
0 & 0 & 0 & \omega_2 \\
0 & 0 & -\omega_2 & 0
\end{pmatrix}, \quad
E_{13} =
\begin{pmatrix}
0 & 0 & 1 & 0 \\
0 & 0 & 0 & 0 \\
0 & 0 & 0 & 0 \\
0 & 0 & 0 & 0
\end{pmatrix}.
\]
We obtain:

{\small
\beqq
e^{\mathbf{A}x}
= \left(
\begin{array}{cccc}
 e^{-\lambda x} \cos (x \omega_1 ) & e^{-\lambda x} \sin (x \omega_1 ) & \frac{\ds \epsilon  e^{-\lambda x}
   (\omega_2  \sin (\omega_2 x)-\omega_1  \sin (x \omega_1 ))}{\omega_2 ^2-\omega_1 ^2} &
   -\frac{\ds \omega_2 \epsilon  e^{-\lambda x} (\cos (\omega_2  x)-\cos (x \omega_1 ))}{\omega_2
   ^2-\omega_1 ^2} \\
 \ds-e^{-\lambda x} \sin (x \omega_1) & e^{-\lambda x} \cos (x \omega_1 ) & -\frac{\ds \omega_1 \epsilon
   e^{-\lambda x} (\cos (\omega_2  x)-\cos (x \omega_1 ))}{\omega_1 ^2-\omega_2^2} &
   \frac{\ds \epsilon  e^{-\lambda x} (\omega_1 \sin(\omega_2 x)-\omega_2 \sin(x \omega_1
   ))}{\omega_2 ^2-\omega_1 ^2} \\
 0 & 0 & e^{-\lambda x} \cos(\omega_2 x) & e^{-\lambda x} \sin(\omega_2 x) \\
 0 & 0 & \ds -e^{-\lambda x} \sin (\omega_2 x) & e^{-\lambda x} \cos(\omega_2 x) \\
\end{array}
\right).
\eeqq
}
We can now compute
\beq
e^{\mathbf{A}^Tx}=(e^{\mathbf{A}x})^T
\eeq
leading to
\begin{multline*}
(e^{\mathbf{A}x}e^{\mathbf{A}^Ty})_{[1,1]}=\frac{\epsilon ^2 e^{-\lambda x-\lambda y} (\omega_2 \sin (\omega_2 x)-\omega_1 \sin (x \omega_1)) (\omega_2  \sin (\omega_2  y)-\omega_1  \sin (y \omega_1 ))}{\left(\omega_2 ^2-\omega_1 ^2\right)^2} + \\
\frac{\omega_2 ^2 \epsilon ^2 e^{-\lambda x-\lambda y} (\cos (\omega_2  x)-\cos (x \omega_1)) (\cos (\omega_2 y)-\cos (y \omega_1 ))}{\left(\omega_2 ^2-\omega_1^2\right)^2} + \\
e^{-\lambda x-\lambda y} (\sin(x \omega_1) \sin(y \omega_1) +  \cos(x \omega_1) \cos(y \omega_1)).
\end{multline*}
\paragraph{Autocorrelation Function.}
The autocorrelation matrix is given by:
\beq
C(t, s) = 2 \sqrt{\sigma_1 \sigma_2} \int_{-\infty}^{\min(t, s)} \exp(A(t - t')) \exp(A^T(s - t')) \, dt'.
\eeq
We focus on the $(1,1)$-component of this matrix, corresponding to the autocorrelation function $C_{xx}(\tau)$ of the variable $X_1(t)$, where $\tau = t - s$, we get the first component of the auto-correlation function
\beqq
C_{xx}(\tau) = 2 \sqrt{\sigma_1\sigma_2} \int_{-\infty}^{s}(e^{\mathbf{A}(t-t')}e^{\mathbf{A}^T(s-t')})_{[1,1]}\mathrm{d}t'.
\eeqq
Finally,
\beqq
C_{xx}(\tau)= Z_{\epsilon}(\omega_1,\omega_2) e^{-\lambda |\tau|} ( &&A_{\epsilon}(\omega_1,\omega_2) \sin(\omega_2 |\tau|) + B_{\epsilon}(\omega_1,\omega_2)\cos(\omega_2\tau) \\&& \\&+& C_{\epsilon}(\omega_1,\omega_2)\sin(\omega_1|\tau|) + D_{\epsilon}(\omega_1,\omega_2)\cos(\omega_1\tau) )
\eeqq
where
\beqq
Z_{\epsilon}(\omega_1,\omega_2)&=& \frac{\sqrt{\sigma_1\sigma_2} }{2 \lambda
(\omega_2^2 -\omega_1^2)\left(\lambda^2+\omega_1 ^2\right) \left((\omega_2
-\omega_1 )^2+4 \lambda^2\right) \left((\omega_2 +\omega_1 )^2+4 \lambda^2\right)}\\ \\
A_{\epsilon}(\omega_1,\omega_2) &=& 4 \omega_2 \lambda \epsilon^2 \left(\lambda^2+\omega_1^2\right) \left(\omega_2^2+4 \lambda^2+\omega_1 ^2\right) \\ \\
B_{\epsilon}(\omega_1,\omega_2) &=& 2 \omega_2^2 \epsilon^2 \left(\lambda^2+\omega_1^2\right) \left(\omega_2^2+4 \lambda^2-\omega_1^2\right)\\ \\
C_{\epsilon}(\omega_1,\omega_2) &=& -\lambda \omega_1 \epsilon^2 \left(\omega_2^4+6 \omega_2^2 \omega_1^2+16 \lambda^4+8 \lambda^2
\left(2 \omega_2^2+\omega_1^2\right)+\omega_1^4\right)
\eeqq
and,
\begin{multline*}
D_{\epsilon}(\omega_1,\omega_2) = -\omega_1^6 (-6 \omega_2^2+18 \lambda^2+\epsilon^2)+2 \omega_2^2 \lambda^2 (\omega_2^2+4 \lambda^2) (\omega_2^2+4 \lambda^2+\epsilon ^2) -2 \omega_1^4(3 \omega_2^4+24 \lambda^4+\lambda^2 (4 \epsilon^2-3 \omega_2^2)) \\ \\+ \omega_1^2 (\omega_2^4 (2 \omega_2^2+\epsilon^2)- 32 \lambda^6+16 \lambda^4 (2 \omega_2^2-\epsilon^2)+2 \lambda^2 (5 \omega_2^4-\omega_2^2 \epsilon^2))-2 \omega_1^8.
\end{multline*}
\paragraph{Power Spectral Density (PSD).}
The PSD is computed as the Fourier transform of $C_{xx}(\tau)$ leading to
\begin{equation}
S(f) = \frac{Z_\epsilon}{2\pi} \left( A_\epsilon F_1(f; \omega_2) + B_\epsilon F_2(f; \omega_2) + C_\epsilon F_3(f; \omega_1) + D_\epsilon F_4(f; \omega_1) \right),
\end{equation}
with spectral kernel functions defined by:
\begin{align*}
F_1(f, \omega) &= \frac{\omega - 2\pi f}{\lambda^2 + (\omega - 2\pi f)^2} + \frac{\omega + 2\pi f}{\lambda^2 + (\omega + 2\pi f)^2}, \\
F_2(f, \omega) &= \frac{\lambda}{\lambda^2 + (\omega - 2\pi f)^2} + \frac{\lambda}{\lambda^2 + (\omega + 2\pi f)^2}, \\
F_3(f, \omega) &= F_1(f, \omega), \\
F_4(f, \omega) &= \frac{\lambda^2 - (\omega - 2\pi f)^2}{(\lambda^2 + (\omega - 2\pi f)^2)^2} + \frac{\lambda^2 - (\omega + 2\pi f)^2}{(\lambda^2 + (\omega + 2\pi f)^2)^2}.
\end{align*}
\paragraph{Total Energy.}
The total energy of the variable $X_1(t)$ is the integral of the PSD:
\begin{equation}
E_{X_1} = \int_{-\infty}^{+\infty} S(f) \, df.
\end{equation}
In the generic case $\omega_1 \ne \omega_2$, the closed-form expression is:
\begin{equation}
E_{X_1} = \frac{2 \sqrt{\sigma_1 \sigma_2}}{16 \lambda \pi} \left( \frac{2 \epsilon^2 (3 \omega_1^2 \omega_2^2 + 16 \lambda^4 + 4 \lambda^2 (\omega_2^2 + 2 \omega_1^2) + \omega_1^4)}{(\lambda^2 + \omega_1^2)[(\omega_2 - \omega_1)^2 + 4 \lambda^2][(\omega_2 + \omega_1)^2 + 4 \lambda^2]} + 4 \right).
\end{equation}
\paragraph{Resonant Case ($\omega_1 = \omega_2 = \omega$).}  
In the special case where both oscillators are tuned to the same frequency, resonance occurs, and the energy diverges in the limit of vanishing damping. A separate expression is given by:
\begin{multline}
C_{xx}(\tau) = \frac{2 \sqrt{\sigma_1 \sigma_2} e^{-\lambda|\tau|}}{16 \lambda^3 \omega (\lambda^2 + \omega^2)^2} \bigg[ \lambda^2 \epsilon^2 (2 \lambda^3 - \lambda^2 \omega^2 |\tau| - \omega^4 |\tau|) \sin(\omega |\tau|) + \\
\omega \cos(\omega \tau) \left(8 \lambda^6 + 2 \lambda^4 \epsilon^2 (2 + \lambda |\tau|) + \lambda^2 \omega^2 (16 \lambda^2 + 3 \epsilon^2(1 + \lambda |\tau|)) + \omega^4 (8 \lambda^2 + \epsilon^2(1 + \lambda |\tau|))\right)\bigg].
\end{multline}
The corresponding PSD is given by
\beq
S(f)= \frac{z}{2\pi}\left(a(\omega,\lambda,\eps)  F_1(f) + b(\omega,\lambda,\eps) F_2(f) + c(\omega,\lambda,\eps) F_3(f) + d(\omega,\lambda,\eps) F_4(f)\right),
\eeq
where
\beqq
F_1(f) &=& \frac{\omega-2\pi f}{\lambda^2+(\omega-2\pi f)^2}+\frac{\omega+2\pi f}{\lambda^2+(\omega+2\pi f)^2} \\
F_2(f) &=& \frac{2\lambda(\omega-2\pi f)}{(\lambda^2+(\omega-2\pi f)^2)^2}+\frac{2\lambda(\omega+2\pi f)}{(\lambda^2+(\omega+2\pi f)^2)^2} \\
F_3(f) &=& \frac{\lambda}{\lambda^2+(\omega-2\pi f)^2}+\frac{\lambda}{\lambda^2+(\omega+2\pi f)^2} \\
F_4(f) &=& \frac{\lambda^2-(\omega-2\pi f)^2}{(\lambda^2+(\omega-2\pi f)^2)^2}+\frac{\lambda^2-(\omega+2\pi f)^2}{(\lambda^2+(\omega+2\pi f)^2)^2}
\eeqq
and
\beqq
a(\omega,\lambda,\eps) &=& 2\lambda^5\epsilon^2\\ \\
b(\omega,\lambda,\eps) &=& -\lambda^2\epsilon^2\omega^2(\lambda^2+\omega^2)\\ \\
c(\omega,\lambda,\eps) &=& \omega(8\lambda^6+4\lambda^4\epsilon^2+\omega^4(8\lambda^2+\epsilon^2)+\lambda^2\omega^2(16\lambda^2+3\epsilon^2))\\ \\
d(\omega,\lambda,\eps) &=& \omega(2\lambda^5\epsilon^2+\omega^4\lambda\epsilon^2+3\lambda^3\omega^2\epsilon^2).\\ \\
z(\omega,\lambda,\eps) &=& \frac{2\sqrt{\sigma_1\sigma_2}}{16 \lambda^3 \omega  \left(\lambda^2+\omega ^2\right)^2}
\eeqq
The total energy is given by
\beq
E_{X_1} &=& \int_{-\infty}^{+\infty}S(f)\mathrm{d}f \nonumber \\ 
  &=& \frac{1}{2\pi} \frac{2\sqrt{\sigma_1\sigma_2}}{16 \lambda^3 \omega \left(\lambda^2+\omega ^2\right)^2} \left(\omega(8\lambda^6+4\lambda^4\epsilon^2+\omega^4(8\lambda^2+\epsilon^2)+\lambda^2\omega^2(16\lambda^2+3\epsilon^2))\right).
\eeq
\paragraph{Conclusion.}
The autocorrelation function and PSD of the first oscillator are analytically tractable for both general and resonant regimes. In the resonant case $\omega_1 = \omega_2$, the effect of coupling $\epsilon$ is amplified, leading to enhanced energy transfer and spectral power near the shared frequency $\omega$. This highlights the importance of frequency detuning and damping in controlling burst synchrony and energy distribution in weakly coupled stochastic oscillators. The analytical expression derived above are used to fit the PSD generated by the simulations of the coupled OU-systems.

\normalem
\bibliographystyle{ieeetr}
\bibliography{biblio,biblioshort}

@article{holcman2006emergence,
	author={Holcman, David AND Tsodyks, Misha},
	journal={PLoS Computational Biology},
	title={The Emergence of Up and Down States in Cortical Networks},
	year={2006},
	month={March},
	volume={2},
	number={3},
	pages={174-181},
	notes={(D)}
}

@book{buzsaki2006rhythms,
  title={Rhythms of the Brain},
  author={Buzsaki, Gyorgy},
  year={2006},
  publisher={Oxford University Press}
}

@article{destexhe2000,
	title={Modelling corticothalamic feedback and the gating of the thalamus by the cerebral cortex},
	author={Destexhe, Alain},
	journal={Journal of Physiology-Paris},
	volume={94},
	number={5-6},
	pages={391--410},
	year={2000},
	publisher={Elsevier}
}

@article{zonca2021modeling,
  title={Modeling bursting in neuronal networks using facilitation-depression and afterhyperpolarization},
  author={Zonca, Lou and Holcman, David},
  journal={Communications in Nonlinear Science and Numerical Simulation},
  volume={94},
  pages={105555},
  year={2021},
  publisher={Elsevier}
}

@article{zonca2021escape,
  title={Escape from an attractor generated by recurrent exit},
  author={Zonca, Lou and Holcman, David},
  journal={Physical Review Research},
  volume={3},
  number={2},
  pages={023115},
  year={2021},
  publisher={APS}
}

@article{verechtchaguina2004spectra,
  title={Spectra and waiting-time densities in firing resonant and nonresonant neurons},
  author={Verechtchaguina, T and Schimansky-Geier, L and Sokolov, IM},
  journal={Physical Review E—Statistical, Nonlinear, and Soft Matter Physics},
  volume={70},
  number={3},
  pages={031916},
  year={2004},
  publisher={APS}
}

@article{verechtchaguina2007interspike,
  title={Interspike interval densities of resonate and fire neurons},
  author={Verechtchaguina, T and Sokolov, Igor M and Schimansky-Geier, Lutz},
  journal={Biosystems},
  volume={89},
  number={1-3},
  pages={63--68},
  year={2007},
  publisher={Elsevier}
}

@book{schuss2009theory,
  title={Theory and applications of stochastic processes: an analytical approach},
  author={Schuss, Zeev},
  volume={170},
  year={2009},
  publisher={Springer Science \& Business Media}
}

@book{gardiner1985handbook,
  title={Handbook of stochastic methods},
  author={Gardiner, Crispin W and others},
  volume={3},
  year={1985},
  publisher={springer Berlin}
}

@incollection{risken1996fokker,
  title={Fokker-planck equation},
  author={Risken, Hannes},
  booktitle={The Fokker-Planck Equation},
  pages={63--95},
  year={1996},
  publisher={Springer}
}

@book{jaffard2001wavelets,
  title={Wavelets: tools for science and technology},
  author={Jaffard, St{\'e}phane and Meyer, Yves and Ryan, Robert D},
  year={2001},
  publisher={SIAM}
}

@article{flandrin2004empirical,
  title={Empirical mode decomposition as a filter bank},
  author={Flandrin, Patrick and Rilling, Gabriel and Goncalves, Paulo},
  journal={IEEE signal processing letters},
  volume={11},
  number={2},
  pages={112--114},
  year={2004},
  publisher={IEEE}
}

@article{flandrin1992wavelet,
  title={Wavelet analysis and synthesis of fractional Brownian motion},
  author={Flandrin, Patrick},
  journal={IEEE Transactions on information theory},
  volume={38},
  number={2},
  pages={910--917},
  year={1992},
  publisher={IEEE}
}

@article{da1973organization,
  title={Organization of thalamic and cortical alpha rhythms: spectra and coherences},
  author={Da Silva, FH Lopes and Van Lierop, THMT and Schrijer, CF and Van Leeuwen, W Storm},
  journal={Electroencephalography and clinical neurophysiology},
  volume={35},
  number={6},
  pages={627--639},
  year={1973},
  publisher={Elsevier}
}

@article{bacsar2012short,
  title={A short review of alpha activity in cognitive processes and in cognitive impairment},
  author={Ba{\c{s}}ar, Erol and G{\"u}ntekin, Bahar},
  journal={International Journal of Psychophysiology},
  volume={86},
  number={1},
  pages={25--38},
  year={2012},
  publisher={Elsevier}
}

@article{chen2014eigenfunctions,
  title={On the eigenfunctions of the complex Ornstein--Uhlenbeck operators},
  author={Chen, Yong and Liu, Yong},
  journal={Kyoto Journal of Mathematics},
  volume={54},
  number={3},
  pages={577--596},
  year={2014},
  publisher={Duke University Press}
}

@book{silverman1972special,
  title={Special functions and their applications},
  author={Silverman, Richard A and others},
  year={1972},
  publisher={Courier Corporation}
}

@article{duc2014oscillatory,
  title={Oscillatory decay of the survival probability of activated diffusion across a limit cycle},
  author={Duc, K Dao and Schuss, Z and Holcman, D},
  journal={Physical Review E},
  volume={89},
  number={3},
  pages={030101},
  year={2014},
  publisher={APS}
}

@article{jacoby2000loess,
  title={Loess:: a nonparametric, graphical tool for depicting relationships between variables},
  author={Jacoby, William G},
  journal={Electoral studies},
  volume={19},
  number={4},
  pages={577--613},
  year={2000},
  publisher={Elsevier}
}

@article{huang1998empirical,
  title={The empirical mode decomposition and the Hilbert spectrum for nonlinear and non-stationary time series analysis},
  author={Huang, Norden E and Shen, Zheng and Long, Steven R and Wu, Manli C and Shih, Hsing H and Zheng, Quanan and Yen, Nai-Chyuan and Tung, Chi Chao and Liu, Henry H},
  journal={Proceedings of the Royal Society of London. Series A: mathematical, physical and engineering sciences},
  volume={454},
  number={1971},
  pages={903--995},
  year={1998},
  publisher={The Royal Society}
}

@article{lopes1974model,
  title={Model of brain rhythmic activity: the alpha-rhythm of the thalamus},
  author={Lopes da Silva, FH and Hoeks, A and Smits, H and Zetterberg, LH},
  journal={Kybernetik},
  volume={15},
  pages={27--37},
  year={1974},
  publisher={Springer}
}

@article{perez2021isostables,
  title={Isostables for stochastic oscillators},
  author={P{\'e}rez-Cervera, Alberto and Lindner, Benjamin and Thomas, Peter J},
  journal={Physical review letters},
  volume={127},
  number={25},
  pages={254101},
  year={2021},
  publisher={APS}
}

@article{thomas2019phase,
  title={Phase descriptions of a multidimensional Ornstein-Uhlenbeck process},
  author={Thomas, Peter J and Lindner, Benjamin},
  journal={Physical Review E},
  volume={99},
  number={6},
  pages={062221},
  year={2019},
  publisher={APS}
}

@book{schuss2011nonlinear,
  title={Nonlinear filtering and optimal phase tracking},
  author={Schuss, Zeev},
  volume={180},
  year={2011},
  publisher={Springer Science \& Business Media}
}

@book{schuss2009diffusion,
	publisher = {Springer-Verlag, New York, NY},
	author = {Schuss, Z.},
	title = {Diffusion and Stochastic Processes. An Analytical Approach},
	year = {2009},
}

@article{bacsar2012review,
  title={A review of alpha activity in integrative brain function: fundamental physiology, sensory coding, cognition and pathology},
  author={Ba{\c{s}}ar, Erol},
  journal={International Journal of Psychophysiology},
  volume={86},
  number={1},
  pages={1--24},
  year={2012},
  publisher={Elsevier}
}

@article{DaoDuc2015,
  title={Bursting reverberation as a multiscale neuronal network process driven by synaptic depression-facilitation},
  author={Dao Duc, K and Lee, Chun-Yao and Parutto, Pierre and Cohen, Dror and Segal, Menahem and Rouach, Nathalie and Holcman, David},
  journal={PloS One},
  volume={10},
  number={5},
  pages={e0124694},
  year={2015},
  publisher={Public Library of Science San Francisco, CA USA}
}

@article{daoduc2016,
  title={Oscillatory survival probability: Analytical and numerical study of a non-Poissonian exit time},
  author={Dao Duc, K and Schuss, Zeev and Holcman, David},
  journal={Multiscale Modeling \& Simulation},
  volume={14},
  number={2},
  pages={772--798},
  year={2016},
  publisher={SIAM}
}

@article{Molle2011,
  author    = {Marc Mölle and Thomas O. Bergmann and Lisa Marshall and Jan Born},
  title     = {Fast and slow spindles during the sleep slow oscillation: disparate coalescence and engagement in memory processing},
  journal   = {Sleep},
  year      = {2011},
  volume    = {34},
  number    = {10},
  pages     = {1411-1421},
  doi       = {10.5665/SLEEP.1290},
}

@article{lindner2004,
	title={Effects of noise in excitable systems},
	author={Lindner, Benjamin and Garc{\i}a-Ojalvo, Jordi and Neiman, Alexander and Schimansky-Geier, Lutz},
	journal={Physics reports},
	volume={392},
	number={6},
	pages={321--424},
	year={2004},
	publisher={Elsevier}
}

@article{linetsky2004computing,
  title={Computing hitting time densities for CIR and OU diffusions: applications to mean-reverting models},
  author={Linetsky, Vadim},
  journal={Journal of Computational Finance},
  volume={7},
  pages={1--22},
  year={2004},
  publisher={Citeseer}
}

@article{davydov2003pricing,
  title={Pricing options on scalar diffusions: an eigenfunction expansion approach},
  author={Davydov, Dmitry and Linetsky, Vadim},
  journal={Operations research},
  volume={51},
  number={2},
  pages={185--209},
  year={2003},
  publisher={INFORMS}
}

@article{linetsky2004spectral,
  title={The spectral decomposition of the option value},
  author={Linetsky, Vadim},
  journal={International Journal of Theoretical and Applied Finance},
  volume={7},
  number={03},
  pages={337--384},
  year={2004},
  publisher={World Scientific}
}

@article{linetsky2004lookback,
  title={Lookback options and diffusion hitting times: A spectral expansion approach},
  author={Linetsky, Vadim},
  journal={Finance and Stochastics},
  volume={8},
  pages={373--398},
  year={2004},
  publisher={Springer}
}

@book{abramowitz1948handbook,
  title={Handbook of mathematical functions with formulas, graphs, and mathematical tables},
  author={Abramowitz, Milton and Stegun, Irene A},
  volume={55},
  year={1948},
  publisher={US Government printing office}
}

@book{bradley1960distribution,
  title={Distribution-free statistical tests},
  author={Bradley, James Vandiver},
  volume={60},
  number={661},
  year={1960},
  publisher={United States Air Force}
}

@book{sheskin2003handbook,
  title={Handbook of parametric and nonparametric statistical procedures},
  author={Sheskin, David J},
  year={2003},
  publisher={Chapman and hall/CRC}
}

@article{verechtchaguina2006_1,
	title={First passage time densities in resonate-and-fire models},
	author={Verechtchaguina, Tatiana and Sokolov, Igor M and Schimansky-Geier, Lutz},
	journal={Physical Review E},
	volume={73},
	number={3},
	pages={031108},
	year={2006},
	publisher={APS}
}

@article{verechtchaguina2006_2,
	title={First passage time densities in non-Markovian models with subthreshold oscillations},
	author={Verechtchaguina, T and Sokolov, IM and Schimansky-Geier, L},
	journal={EPL (Europhysics Letters)},
	volume={73},
	number={5},
	pages={691},
	year={2006},
	publisher={IOP Publishing}
}

@book{bacsar2012brain,
  title={Brain function and oscillations: volume I: brain oscillations. Principles and approaches},
  author={Ba{\c{s}}ar, Erol},
  year={2012},
  publisher={Springer Science \& Business Media}
}
\end{document}